\documentclass[12pt]{article}
\usepackage{draft,float,mathdots}

\begin{document}

\unitlength = .8mm

\begin{titlepage}

\begin{center}

\hfill \\
\hfill \\
\vskip 1cm

\title{Deformations with Maximal Supersymmetries
\\
Part 1: On-shell Formulation}

\author{Chi-Ming Chang$^a$, Ying-Hsuan Lin$^a$, Yifan Wang$^b$, Xi Yin$^{a}$}

\address{$^a$Jefferson Physical Laboratory, Harvard University, \\
Cambridge, MA 02138 USA
\\
$^b$Center for Theoretical Physics, Massachusetts Institute of Technology, \\
Cambridge, MA 02139 USA}

\email{cmchang@physics.harvard.edu, yhlin@physics.harvard.edu, \\ yifanw@mit.edu,
xiyin@fas.harvard.edu}

\end{center}

\abstract{ We study deformations of maximally supersymmetric gauge theories by higher dimensional operators in various spacetime dimensions. We classify infinitesimal deformations that preserve all 16 supersymmetries, while allowing the possibility of breaking either Lorentz or R-symmetry, using an on-shell algebraic method developed by Movshev and Schwarz. We also consider the problem of extending the deformation beyond the first order. }

\vfill

\end{titlepage}

\eject

\tableofcontents

\section{Introduction}

It has long been recognized that supersymmetry puts highly nontrivial constraints on the structure of quantum field theories that \cite{Gates:1983nr, Mandelstam:1982cb, Brink:1982wv, Howe:1983sr}, on one hand allows for exact solutions of certain physically relevant observables \cite{Seiberg:1994rs, Pestun:2007rz}, and on the other hand retains rich and complex dynamics, including those that are responsible for holographic duality with gravity \cite{Maldacena:1997re}. It is often asserted that the greater the number of supercharges, the simpler the quantum field theory would be, and the maximally supersymmetric Yang-Mills theory (MSYM) would be the simplest of them all, thereby dubbed ``the harmonic oscillator of the 21st century" \cite{Arkani-Hamed:interview}. It might then seem odd that no simple\footnote{By simple we mean one that requires introducing only finitely many auxiliary fields.} off-shell superspace formulation exists that makes all 16 supersymmetries manifest \cite{Galperin:2001uw}, and it is not always easy to make non-renormalization arguments that utilize the full power of maximal supersymmetry. Examples where such non-renormalization theorems are desired include the derivative expansion of the effective theory on the Coulomb branch moduli space of MSYM \cite{Dine:1997nq, Seiberg:1997ax, Paban:1998qy, Nicolai:2000ht, Maxfield:2012aw}, and the constraints on loop divergences and counter terms in MSYM in more than 4 dimensions \cite{Howe:2002ui, Berkovits:2009aw} (and the analogous questions in supergravity with 32 supersymmetries \cite{Bern:2009kd, Elvang:2010jv, Beisert:2010jx}). In practice one typically works either with component fields, where supersymmetries are realized on-shell, or invokes arguments based on superspace formalism that makes 8 or fewer supersymmetries manifest.\footnote{Techniques based on on-shell scattering amplitudes have been particularly powerful and useful in 4 dimensions, though these are rather different from the approach taken in the present paper.}

Methods of dealing with maximally supersymmetric gauge theories with all 16 supersymmetries manifest have been developed, both in the on-shell formulation based on the associative algebra of super-gauge covariant derivatives and its deformations, by Movshev and Schwarz \cite{Movshev:2003ib, Movshev:2004aw, Movshev:2005ei, Movshev:2009ba}, and in the off-shell formulation based on pure spinor superspace \cite{Nilsson:1985cm,Tonin:1991ii,Howe:1991mf,Howe:1991bx,Berkovits:2001rb, Cederwall:2011vy, Cederwall:2013vba}. These methods will be heavily employed in our paper.
As a matter of terminology, in this paper we refer to all higher derivative gauge theories based on Abelian or non-Abelian gauge groups as MSYM, or ``deformed" MSYM. The most familiar two-derivative super-Yang-Mills theory will be referred to as the ``undeformed" MSYM. The first question we would like to address in this paper is, what sort of higher derivative deformations of the Lagrangian are allowed by 16 supersymmetries. This is a subtle question in the component field formulation of MSYM, because there the supersymmetry transformations only close on-shell. It is generally necessary to deform the supersymmetry transformations along with the Lagrangian, and it is insufficient to deform the Lagrangian only to first order by an operator of given scaling dimension. See \cite{Bergshoeff:2000cx, Bergshoeff:2001dc, Sevrin:2001ha,Cederwall:2001bt,Cederwall:2001td,Cederwall:2001dx,Collinucci:2002ac, Howe:2010nu,Bossard:2010pk} for works in this direction in the component field formulation.

In the case of single trace deformations in large $N$ gauge theories that respect both Lorentz and R-symmetries, at the level of first order deformations, this problem was solved by  \cite{Movshev:2005ei} in ten dimensions (deformations of classical 10D SYM) and by  \cite{Movshev:2009ba} in zero dimension (IKKT matrix model), via the study of deformations of the associative algebra generated by super-gauge covariant derivatives subject to the equations of motion (for superfields). Identifying obstruction classes and proving their absence for the corresponding higher order deformations are generally difficult. We will examine this problem, for MSYM in all dimensions from zero to ten, and also consider deformations that break either Lorentz or R-symmetries.\footnote{We have in mind the application to for instance the study of Coulomb branch effective actions, though in this example the deformations of interest are not of the single trace type (they are non-polynomial).}

At the level of first order (i.e., infinitesimal) single trace deformations, we present a classification. Such infinitesimal deformations fall into three classes, which we refer as F-term deformations, D-term deformations, and exceptional D-term deformations. If one demands both Lorentz and R-symmetry invariance, then the only single trace F-term deformation is the Born-Infeld deformation, roughly speaking the supersymmetric completion of ${\rm Tr} F^4$ term. In the R-symmetry preserving, Lorentz violating case, the only F-term deformations are those that correspond to noncommutative MSYM theories. In the Lorentz invariant, R-symmetry violating case, the only F-term deformations transform in the symmetric traceless tensor representations of the R-symmetry group $SO(10-d)$, where $d$ is the spacetime dimension, with an exception in the zero dimension case, where there is an additional 5-form deformation in the IKKT matrix model. All of the F-term deformations, at the infinitesimal level, can be realized as a Lagrangian deformation by some number of supercharges acting on a half-BPS operator~\cite{DHoker:2002aw}, and have simple interpretations from the holographic duality perspective. Interestingly, they are not always ``half superspace integrals", in that there can be fewer than 8 supercharges acting on a half-BPS operator and still result in a fully supersymmetric deformation. The D-term deformations are ``full superspace integrals", i.e., constructed from all 16 supercharges acting on a non-BPS operator. These are generic in any MSYM theories. The exceptional D-term deformations are not quite full superspace integrals, in that they can be obtained by taking all 16 supercharges acting on a gauge-non-invariant expression constructed out of the vector potentials and not just the field strengths. These appear only in spacetime dimension 8 and higher.

This on-shell algebraic approach has in principle the advantage that it formulates the problem of finding higher order deformations (or identifying the obstructions) systematically as a cohomology problem. In practice, however, it can be very difficult to compute the relevant obstruction classes, due to the non-explicit nature of theorems that relate certain Hochschild cohomology of interest to the cohomology of a pure spinor complex. In simple cases such as the noncommutative deformations and the 5-form deformation of IKKT matrix model, we can find higher order on-shell deformations by direct computation, but this is hard to do for the Born-Infeld deformation. In a companion paper, we will solve the formal deformation problem for the Born-Infeld term in the off-shell approach based on pure spinor superspace.

In section 2 we review the construction of the associative algebra that captures the equations of motion of MSYM theories, and the reformulation of the deformation problem in terms of certain cohomology groups. We leave many important but technical details to the Appendices, while presenting the result of the classification of infinitesimal deformations in section 3. We discuss the higher order deformations in the on-shell approach in section 4, and conclude in section 5.

\section{The Super-Yang-Mills algebra and its deformations} 
\label{SYM algebra and its deformations}

\subsection{Algebraization of the problem}

We begin with the on-shell superfield formalism of MSYM, and will soon reformulate deformations of the SYM equation of motion in terms of suitable deformations of the associative algebra generated by super-gauge covariant derivatives. For the moment we will adopt 10-dimensional notation, and write the Yang-Mills superfield as $A_\A(x,\theta)$, where $x^m$ are the bosonic spacetime coordinates and $\theta^\A$ fermionic coordinates. We use upper spinor indices to denote the chiral spinor representation of $Spin(10)$\footnote{For simplicity, we will be working in the Euclidean signature.} and lower indices for the anti-chiral spinor. The gamma matrices acting on chiral or anti-chiral spinors are denoted $\Gamma^m_{\A\B}$ or $(\Gamma^m)^{\A\B}$. $\Gamma^{m_1\cdots m_k}$ denote the antisymmetrized product of gamma matrices as usual. Note that while $\Gamma^m$ and $\Gamma^{mnpqr}$ are symmetric matrices, $\Gamma^{mnp}$ is anti-symmetric. Denote by $d_\A$ the ordinary super-derivative
\ie
d_\A = {\partial\over \partial \theta^\A} + {1\over 2} (\Gamma^m\theta)_\A {\partial\over \partial x^m},
\fe
and by $D_\A$ the the gauge covariant super-derivative,
\ie
D_\A = d_\A + A_\A.
\fe
The undeformed SYM equation of motion is equivalent to the quadratic relation on $D_\A$,
\ie\label{LCT}
(\Gamma^{mnpqr})^{\A\B}\{D_\A,D_\B\}=0.
\fe
%In this subsection, we introduce the Lie algebras $L$ and $YM$, and reformulate the problem of classifying the deformations of the $d$-dimensional maximally supersymmetric Yang-Mills theory (MSYM$_d$) as the problem of classifying deformations of the algebra $U(YM)$.
This is equivalent to the statement that
\ie
\{D_\A,D_\B\}=\Gamma^m_{\A\B}D_m,
\fe
for {\it some} operator $D_m$ (which may be defined as the gauge covariant bosonic derivative).

We now view (\ref{LCT}) as the defining relation on the generators $D_\A$ of a graded Lie super-algebra $L$. $D_\A$ are the only level 1 elements of $L$. This is appropriate for $U(N)$ gauge theory in the $N\to \infty$ limit, as there are no further independent relations. The level 2, 3, 4 components of $L$ are spanned by 
\ie
&D_n\equiv{1\over 16}\Gamma^{\A\B}_n\{D_\A,D_\B\},~~~\chi^\A\equiv{1\over 10}\Gamma^{\A\B}_n[D_\B,D_n],~~~F_{mn}\equiv[D_m,D_n].
\fe
%By the constraint of the algebra \eqref{LCT}, $D_n$ are the only independent grade-2 elements, i.e.
When $D_\A$ is expressed in terms of a superfield $A_\A(x,\theta)$, $D_m,\chi^\A,F_{mn}$ have the interpretation as the bosonic super-covariant derivative, the gaugino, and the field strength superfields. It is easy to show that
\ie\label{SV1}
\{D_\A,D_n\}=\Gamma^n_{\A\B}\chi^\B,
%hence $\chi^\A$ are the only independent grade-3 elements. Similarly, $F_{mn}$ are the only grade-4 elements. The grade-4 elements $\{D_\A,\chi^\B\}$ can be written in terms of $F_{mn}$ as
~~~~ \{D_\A,\chi^\B\}={1\over 4}(\Gamma_{mn})_\A{}^\B F_{mn}.
\fe
and thus $\chi^\A$ and $F_{mn}$ are indeed the only independent elements of $L$ at level 3 and 4.
It also follows from their definition and the defining relation on $D_\A$'s that $D_m,\chi^\A,F_{mn}$ obey
\ie\label{MYMEOM}
&\Gamma^n_{\A\B} [D_n,\chi^\B]=0,
\\
&[D_m,F_{mn}]+\Gamma^n_{\A\B}\{\chi^\A,\chi^\B\}=0,
\fe
which takes exactly the same form as the equations of motion of MSYM in component fields, derived from the Lagrangian
\ie
{\cal L}_{SYM}=\tr\left({1\over 4}[D_m, D_n][D_m, D_n]+ \Gamma^n_{\A\B}\chi^\A[D_n,\chi^\B]\right).
\fe
%The (anti-)commutator of $D_\A$ with $D_m,\chi^\A$, \eqref{SV1} and \eqref{SV2}, can be interpreted as supersymmetric variation on the covariant and the gagino field. 
Later we will consider deformations of MSYM equations of motion. Instead of working with the Lagrangian or the component field form of the equations, we will think of these deformations as deforming the algebraic relation of $D_\A$'s, to be described more precisely below. %More precisely, we will be considering deformations of the Lie bracket of $L$ that take value in the associative algebra $U(YM_d)$ which is also an $L$-module.

Denote by $L^i$ the level $i$ component of $L$. We can split $L$ according to its grading,
\ie
L = \bigoplus_{i=1}^\infty L^i.
\fe
It will be useful to define the following graded Lie subalgebras of $L$,
\ie\label{ymdx}
YM_d \equiv \left\langle \Phi_{d+1} ,\cdots, \Phi_{10} \right\rangle \oplus \bigoplus_{i=3}^\infty L^i,
\fe
where we wrote $\Phi_m\equiv D_m$ for $m=d+1,\cdots, 10$, corresponding to the scalar fields in the reduction of 10D SYM to $d$ dimensions. In the notation of \cite{Movshev:2009ba},
\ie
YM \equiv YM_0 = \bigoplus_{i=2}^\infty L^i,~~~~ TYM \equiv YM_{10} = \bigoplus_{i=3}^\infty L^i.
\fe
$YM$ may also be defined as the Lie algebra generated by the level 2 even elements $D_n$ and the level 3 odd elements $\chi^\A$, with the relations \eqref{MYMEOM}. This is because (anti-)commutators of level 3 and higher elements with $D_\A$ can always be rewritten as commutators with $D_m$. Note that $TYM$ is in fact a free Lie algebra generated by $D_m$-derivatives of $\chi^\A$ and $F_{mn}$.
We will often make use of the universal enveloping algebras of $L$ and $YM_d$, which will be denoted by $U(L)$ and $U(YM_d)$ respectively. 

%In this paper, we entirely focus on the MSYM with gauge group $U(\infty)$, such that there are no other relations among the gauge covariant operators than the equation of motion \eqref{MYMEOM}. 
For $U(N)$ gauge theory in the $N\to \infty$ limit, the classical equations of motion are completely encoded in the relations of $U(YM)$. There is a one-to-one correspondence between consistent deformations of the MSYM equations of motion in $d$ spacetime dimensions and deformations of the Lie bracket of $YM$ that take value in a correspondingly deformed version of the associative algebra $U(YM_d)$, that is compatible with the Jacobi identity of the Lie bracket. At the infinitesimal level, this is classified by the Lie algebra cohomology ${\rm H}^2(YM,U(YM_d))$.\footnote{For the infinitesimal deformations, the associative algebra structure on $U(YM_d)$ is not needed, and it suffices to regard $U(YM_d)$ as a $YM$-module, which is isomorphic to ${\rm Sym}(YM_d)$, the direct sum of all symmetric tensor powers of $YM_d$.} Some basic notions and results of the deformation theory of Lie algebras and associative algebras are reviewed in Appendix \ref{Lie algebra cohomology}. %In particular, infinitesimal deformations of the universal enveloping algebra $U(\cG)$ of a Lie algebra $\cG$ is classified by the Lie algebra cohomology ${\rm H}^2(\cG,U(\cG))$.\footnote{Generally, infinitesimal deformations of an associative algebra $A$ is classified by the second Hochschild cohomology ${\rm HH}^2(A,A)$. In the special case $A=U({\cal G})$ for some Lie algebra ${\cal G}$, ${\rm HH}^2(A,A)$ is the same as the Lie algebra cohomology ${\rm H}^2({\cal G}, U(\cG))$ via the Cartan-Eilenberg isomorphism.} 

%In our case, this implies that the infinitesimal equation of motion deformation of MSYM$_0$ is classified by the cohomology group ${\rm H}^2(YM,U(YM))$. 

We are interested in supersymmetric deformations. %They can be characterized as deformations of the Lie bracket on $YM$, valued in an associative algebra deformed from $U(YM_d)$, that are compatible with an action of $D_\A$ on the fields (the supersymmetry transformation of the fields is generally deformed as well). %In the $d=0$ example, namely the IKKT matrix model, 
It is explained in Section \ref{formald} that the { infinitesimal} (i.e. first order) deformations of superfield equations of motion are classified by the cohomology group ${\rm H}^2(L,U(YM_d))$. They would induce supersymmetric deformations on the equations of motion of component fields, which are classified by the image of
\ie
i^*:{\rm H}^2(L,U(YM_d))\to{\rm H}^2(YM,U(YM_d)).
\fe
Here $i^*$ is the map induced by the inclusion $i:YM\hookrightarrow L$, with $YM$ viewed as an ideal of $L$, and is analyzed in Appendix~\ref{Kernel of i^*}.

Once we have identified an infinitesimal supersymmetric deformation as a cohomology class in ${\rm H}^2(L, U(YM_d))$, we may ask whether it can be extended to a formal deformation to all orders. There is a systematic procedure of identifying the obstruction class at every order, which lies in ${\rm H}^3(L,U(YM_d))$, via Gerstenhaber brackets \cite{Doubek}. If the $n$-th order obstruction class is trivial in ${\rm H}^3(L,U(YM_d))$, then there is a coboundary representative that can be used to determine the $n$-th order deformation of the Lie bracket on $L$. Note that the higher order deformations generally do not correspond to cohomology classes in ${\rm H}^2(L,U(YM_d))$. This construction is a slight generalization of the formal deformation theory of an associative algebra, which is reviewed in Appendix \ref{formal}.

\subsection{Identifying all infinitesimal deformations}
\label{sec:inf-def}

In order to classify infinitesimal supersymmetric deformations, we need to identify elements of the the cohomology ${\rm H}^2(L, U(YM_d))={\rm H}^2(L, {\rm Sym}(YM_d))$.\footnote{By Poincar\'e-Birkhoff-Witt theorem~\cite{PBW, PBW-W}, we can replace $U(YM_d)$ by the direct sum of all symmetric tensor powers of $YM_d$. Each symmetric power is independently an $L$-module. } In this subsection, we describe the logic in this computation, leaving many details to the appendices. A key result of~\cite{polishchuk2005quadratic, Movshev:2009ba}, proven based on quadratic duality of Koszul algebras, is the isomorphism (reviewed in Appendix~\ref{koszul})
\ie\label{psf}
{\rm H}^*(L, {\rm Sym}(YM_d)) \simeq {\rm H}^*({\rm Sym}(YM_d)\otimes {\cal S}, Q=\lambda^\A D_\A).
\fe
Here ${\cal S}$ is the ring of polynomials in pure spinor variables $\lambda^\A$. Namely, $\lambda^\A$ is a complex spinor variable subject to the quadratic constraint $\lambda^\A \Gamma^m_{\A\B}\lambda^\B = 0$. ${\rm Sym}(YM_d)\otimes {\cal S}$ is decomposed into a cochain complex according to the grading, with the coboundary operator given by $d=\lambda^\A D_\A$, where $\lambda^\A$ acts on ${\cal S}$ by multiplication and $D_\A$ acts on ${\rm Sym}(YM_d)$ by (anti-)commutators.

It is easy to understand how to go between a cohomology class in ${\rm H}^2({\rm Sym}(YM_d)\otimes {\cal S})$ and an infinitesimal deformation of the superfield equations of motion. The former is represented by a cocycle of the form $\lambda^\A \lambda^\B {\cal O}_{\A\B}$, ${\cal O}_{\A\B}\in{\rm Sym}(YM_d)$. The corresponding deformation of the MSYM equation of motion is
\ie
\{ D_\A, D_\B\} = \Gamma^m_{\A\B} D_m + \epsilon {\cal O}_{\A\B} + {\cal O}(\epsilon^2).
\fe
Indeed the cocycle condition on ${\cal O}_{\A\B}$ simply follows from the Jacobi identity on the nested commutator of $D_\A$'s to first order in $\epsilon$. %See Appendix \ref{Supersymmetric deformations} for more details.

The cohomology groups on the RHS of (\ref{psf}) is then computed by geometric representation theory techniques. First, one ``lifts" the cochain complex of vector spaces ${\rm Sym}(YM_d)\otimes {\cal S}$ to a cochain complex of vector bundles over the projective pure spinor space ${\cal Q}$ (see Appendix~\ref{sec:bundles}), replacing the degree $k$ component ${\cal S}_k$ by the line bundle ${\cal O}(k)$ over ${\cal Q}$.  This complex of vector bundles may be expressed as a direct sum of symmetric tensor powers, ${\rm Sym}({\cal YM}_d)$, where ${\cal YM}_d$ is the complex $\bigoplus_k YM_d\otimes {\cal O}(k)$. The differential $Q=\lambda^\A D_\A$ naturally lifts to a coboundary operator acting on the sections of the bundle ${\rm Sym} ({\cal YM}_d)$,
\ie
Q: \Omega^a({\rm Sym}(YM_d)\otimes {\cal O}(k)) \to \Omega^a({\rm Sym}(YM_d)\otimes {\cal O}(k+1)),
\fe
simply by regarding $\lambda^\A$ as a section of ${\cal O}(1)$. Together with the Dolbeault operator $\bar\partial: \Omega^a\to \Omega^{a+1}$, one obtains a double complex of sections of vector bundles over ${\cal Q}$.  

The idea here is that the cohomology groups in (\ref{psf}) are related to the hypercohomology of this complex of vectors bundles, namely the cohomology of the diagonal differential $\bar\partial + Q$ on the above-mentioned double complex. The latter is computable thanks to the fact that, on a given fiber over ${\cal Q}$, the cohomology of $Q=\lambda^\A D_\A$ (now $\lambda^\A$ regarded as a fixed pure spinor) is very simple. Furthermore, there is a quasi-isomorphism between ${\cal YM}_d$ and a two-term complex $( (L^2)_d \to {\cal W} ) \otimes \cO(2)$ of vector bundles over ${\cal Q}$.  This allows us to collapse the complex of vector bundles to a two-term complex, whose hypercohomology can then be deduced using spectral sequence techniques and Borel-Weil-Bott theorem. The details of this computation are explained in Appendix~\ref{Hypercohomology}.

The relation between (\ref{psf}) and the hypercohomology is understood through a spectral sequence argument sketched below. If we first take the cohomology of the double complex with respect to $\bar\partial$,  and use the fact that the only non-vanishing Dolbeault cohomology groups of the line bundle ${\cal O}(k) \to {\cal Q}$ are
\ie
{\rm H}^0({\cal Q}, {\cal O}(k)) \simeq {\cal S}_k~~~(k\geq 0),~~~~ {\rm H}^{10}({\cal Q}, {\cal O}(k))\simeq {\cal S}^*_{-8-k}~~~~(k\leq -8),
\fe
then the differential $Q$ of the double complex induces a coboundary operator on the $\bar\partial$-cohomology, which is closely related to the complex ${\rm Sym}(YM_d)\otimes {\cal S}$. More precisely, the cohomology of $Q=\lambda^\A D_\A$ in the complex ${\rm Sym}(YM_d)\otimes {\cal S}$ as well as the dual complex ${\rm Sym}(YM_d)\otimes {\cal S}^*$ appear on the second page of a spectral sequence that converges to the hypercohomology of ${\rm Sym}({\cal YM}_d)$. 
Inspection of this spectral sequence results in a long exact sequence
\ie
&\cdots \to {\rm H}_1(L,{\rm Sym} (YM_d))_{\ell-8} \to^{\!\!\!\!\!\!\delta}~ {\rm H}^2(L,{\rm Sym} (YM_d))_\ell \to {\bf H}^{2}({\cal Q},{\rm Sym} ({\cal YM}_d))_\ell
\\
&~~~ \to^{\!\!\!\!\!\!\iota}~{\rm H}_0(L,{\rm Sym} (YM))_{\ell-8} \to {\rm H}^3(L,{\rm Sym} (YM))_\ell\to\cdots
\label{les2}
\fe
Here ${\bf H}^*({\cal Q}, {\rm Sym}({\cal YM}_d))$ stands for the hypercohomology of the double complex $\Omega^*({\rm Sym}({\cal YM}_d))$. The subscript $\ell$ indicates the grading.  Details of this derivation can be found in Appendix~\ref{sec:les}.

The cohomology group of interest is ${\rm H}^2(L,{\rm Sym} (YM_d))$ (recall that its image under $i^*$ in ${\rm H}^2(YM,{\rm Sym} (YM_d))$ classifies fully supersymmetric deformations). The cokernel of $\delta$ in (\ref{les2}) can be identified within the hypercohomology ${\bf H}^2({\cal Q}, {\rm Sym}({\cal YM}_d))$, which is computed explicitly in Appendix~\ref{Hypercohomology}. Loosely speaking, $\delta$ plays the role of an integration over the full superspace. The elements in the cokernel of $\delta$, or equivalently the kernel of $\iota$, will be identified as F-term deformations.

The image of $\delta$ in ${\rm H}^2(L,{\rm Sym} (YM_d))$, on the other hand, fits in the following commutative diagram,
\begin{diagram}\label{commdiag}
{\bf H}^1({\cal Q},{\rm Sym}({\cal YM}_d))_\ell &  \rTo^{\iota} & {\rm H}_{1}(L,{\rm Sym}(YM_d))_{\ell-8} & \rTo^\delta & {\rm H}^{2}(L,{\rm Sym}(YM_d))_\ell & \rTo & {\bf H}^2({\cal Q},{\rm Sym}({\cal YM}_d))_\ell
\\
 &\ruTo(2,2)^{i_*}  & &&\dTo(2,2)^{i^*}
\\
{\rm H}_{1}(YM,{\rm Sym}(YM_d))_{\ell-8} & \rTo^{A_1} & {\rm H}_{1}(YM,{\rm Sym}(YM_d))_{\ell+8} & \rTo^{P}_\cong & {\rm H}^{2}(YM,{\rm Sym}(YM_d))_\ell
\\
\uTo^{B_{YM}} & & \uTo^{B_{YM}}  
\\
{\rm H}_{0}(YM,{\rm Sym}(YM_d))_{\ell-8}& \rTo^{A_0} &{\rm H}_{0}(YM,{\rm Sym}(YM_d))_{\ell+8}&
\end{diagram}

\noindent Here $i_*$ and $i^*$ are respectively the maps on the Lie algebra homology and cohomology induced by the inclusion $YM\hookrightarrow L$. Recall that it is really the image of $i^*$ that gives nontrivial supersymmetric deformations. The map $B_{YM}: {\rm H}_0(YM, {\rm Sym}(YM_d)) \to {\rm H}_1(YM, {\rm Sym}(YM_d))$ is the Connes differential \cite{LodayQuillen}, which amounts to varying a deformation term in the Lagrangian.\footnote{
An alternative definition of the Connes differential $B_{YM}: {\rm H}_0(YM, {\rm Sym}(YM_d)) \to {\rm H}_1(YM, {\rm Sym}(YM_d))$ without reference to cyclic homology can be found in Appendix~\ref{susyhom}.  The Connes differential on higher homology groups is not needed.
}
The map $A_0: {\rm H}_0(YM, {\rm Sym} (YM_d)) \to {\rm H}_0(YM, {\rm Sym} (YM_d))$ amounts to performing a full superspace integral. Namely, it takes ${\rm tr} (G) \in {\rm H}_0(YM, {\rm Sym} (YM_d))$ to $\epsilon^{\A_1\cdots\A_{16}} D_{\A_1}\cdots D_{\A_{16}}{\rm tr}(G)$. The map $A_1$ may be defined in a similar manner on representatives of ${\rm H}_1(YM, {\rm Sym} (YM_d))$. The map $P:{\rm H}_1(YM, {\rm Sym} (YM_d) )\to {\rm H}^2(YM, {\rm Sym} (YM_d))$ is a Poincar\'e isomorphism, whose existence is a nontrivial property of the Lie super-algebra $YM$, and is proven by \cite{Movshev:2004aw} and reviewed in Appendix~\ref{Poincare isomorphism}. It amounts to converting a D-term deformation in the equations of motion for component fields to a deformation of the superfield equations.
%Details about the Connes differentials $B_{YM}, B_L$ can be found in Appendix~\ref{susyhom}.

Now the image of $\delta$ that comes from ${\rm Im} ( i_*\circ B_{YM})\subset {\rm H}_1(L,U(YM_d))$ are identified with the D-term deformations, whereas the image of $\delta$ coming from the cokernel of $i_*\circ B_{YM}$ will be referred to as exceptional D-term deformations. %{\bf Why are the exceptional D-term deformations of the $A{\rm tr}(\widetilde G)$ form?}
The exceptional D-term deformations can be studied via the following commutative diagram,
\begin{diagram}\label{commdiag2}
 {\rm H}_{0}(L,{\rm Sym}(YM_d))_{\ell} & \rTo^{B_{L}} & {\rm H}_{1}(L,{\rm Sym}(YM_d))_{\ell}&  \lTo^{\iota}& {\bf H}^1({\cal Q},{\rm Sym}({\cal YM}_d))_\ell 
\\
\uTo^{i_*} & & \uTo^{i_*}  
\\
{\rm H}_{0}(YM,{\rm Sym}(YM_d))_{\ell}& \rTo^{B_{YM}} &{\rm H}_{1}(YM,{\rm Sym}(YM_d))_{\ell}& 
\end{diagram}
where $B_L:{\rm H}_0(L, {\rm Sym}(YM_d)) \to {\rm H}_1(L, {\rm Sym}(YM_d))$ is the Connes differential. Since the left $i_*$ is obviously surjective, it follows that the cokernel of $i_*\circ B_{YM}$ is the same as the cokernel of $B_L$. Hence, the exceptional D-term deformations are classified by ${\rm coker}(B_L)$ modulo the image of $\iota$. The cokernel of $B_L$ can be studied using the spectral sequence 
\ie\label{SHSS}
{\rm E}^{i,j}_1={\rm H}_{i-j}(L, {\rm Sym}^j(YM_d))\Rightarrow {\rm H}_{i+j}(L/YM_d,\bC)={\rm H}_{i+j}({\bf susy}_d,\bC),
\fe

\noindent where ${\bf susy}_d= L/YM_d$ is the supersymmetry algebra in $d$ spacetime dimensions. The differential $d_0$ on the zeroth page is given by the boundary map for Lie algebra homology $d:\Lambda^{i-j}L\otimes {\rm Sym}^j(YM_d)\to \Lambda^{i-j-1}L\otimes {\rm Sym}^j(YM_d)$. The differential $d_1$ on the first page is a map induced by the inclusion $YM_d \hookrightarrow L$; in other words, $d_1$ is the Connes differential $B_L$. As is proven in \cite{Movshev:2005ei}, the spectral sequence \eqref{SHSS} stabilizes on the second page (elucidated in Appendix~\ref{susyhom}); furthermore, the image of $\iota$ inside ${\rm E}^{i,j}_1$ stabilizes on the first page. Hence, the cokernel of $B_L$ is identical to the SUSY homology ${\rm H}_{i+j}({\bf susy}_d,\bC)$, and the exceptional D-term deformations are classified by ${\rm H}_{i+j}({\bf susy}_d,\bC)$ modulo the image of $\iota$. The homology of ${\bf susy}_d$ is computed in \cite{Movshev:2011pr,Brandt:2009xv,Brandt:2010fa,Brandt:2010tz}, and the results are summarized in Appendix \ref{susyhom}.

\subsection{Formal deformations}
\label{formald}

Starting with an infinitesimal deformation, one can try to construct an all-order formal deformation of the MSYM superfield equations of motion. Such a deformation consists of the following data. We have a formal deformation of the Lie bracket of $L$ taking value in $N=U(YM_d)$, a deformation of the representation $L \to {\rm End}(N)$, and a deformation of the associative algebra multiplication map $N\otimes N\to N$, that obey a set of compatibility conditions. 

Generally, given a Lie algebra $\cG$\footnote{The generalization to Lie superalgebras is straightforward.}, a Lie-ideal ${\cal H}$, and $N=U({\cal H})\subset U(\cG)$ a ${\cal G}$-module by adjoint action, a formal deformation of the Lie bracket together with that of the representation $N$ is described by a skew-symmetric bilinear map 
\ie
\varphi^t = \sum_{n=1}^\infty t^n \varphi_n~:~\Lambda^2\cG\to N,
\fe
together with a representation map 
\ie
\rho^t = \sum_{n=0}^\infty t^n \rho_n~:~ \cG\otimes N\to N,
\fe
with $\rho_0(a,x) = [a,x]$ the undeformed adjoint action of $\cG$ on $N$, and a multiplication map
\ie
m^t = \sum_{n=0}^\infty t^n m_n~:~N\otimes N\to N,
\fe
where $m_0(x,y) = xy$ is the undeformed product in $N$.
They obey the compatibility conditions (here we omit the appropriate signs in dealing with graded Lie super-algebras)
\ie\label{compc}
& \rho^t(a,b)= [a,b]+\varphi^t(a\wedge b),~~~a\in\cG, ~~ b\in \cH=\cG\cap N,
\\
& m^t(a,x)-m^t(x,a)  = \rho^t(a,x),~~~a\in\cH,~~x\in N,
\fe
and the associativity identities (or Jacobi identities)
\ie\label{jac}
& \varphi^t(a\wedge [b,c]) + \rho^t(a,\varphi^t(b\wedge c)) + ({\rm cyclic~permutations}) = 0,
\\
& \rho^t(a, \rho^t(b,x)) - \rho^t(b, \rho^t(a,x)) = \rho^t([a,b],x) + \rho^t(\varphi^t(a\wedge b),x),
\\
& \rho^t(a, m^t(x,y)) = m^t(\rho^t(a,x) ,y) + m^t(x,\rho^t(a,y)),
\\
& m^t(m^t(x,y),z) = m^t(x,m^t(y,z)).
\fe
If $N$ is $U(\cG)$, $\varphi^t$ and $\rho^t$ would be just given in terms of restrictions of $m^t$, and we would be just talking about formal deformations of the associative algebra $U(\cG)$. When $N$ is not the same as $U(\cG)$, we have here a more general notion of a formal deformation, described by the triple $(\varphi^t, \rho^t, m^t)$.  Two deformations $(\varphi^t, \rho^t, m^t)$ and $(\widetilde\varphi^t, \widetilde\rho^t, \widetilde m^t)$ are equivalent if they are related by a pair of ``formal isomorphism maps" $a\mapsto a+ f^t(a)$, $f^t = \sum_{n=1}^\infty t^n f_n \in {\rm Hom}(\cG, N)$, and $h^t=\sum_{n=0}^\infty t^n h_n\in {\rm End}(N)$, with $h_0 = {\rm Id}$, satisfying the compatibility condition $f_n(b)=h_n(b)$ for $b\in\cH=\cG\cap N$.

The equivalence relations on the deformations are
\ie\label{auto}
&  h^t(\widetilde \varphi^t(a\wedge b)) + f^t([a,b]) 
\\
&~~~ = \varphi^t(a\wedge b) 
+ \rho^t(a, f^t(b)) - \rho^t(b, f^t(a))
+ m^t(f^t(a), f^t(b)) - m^t(f^t(b), f^t(a)),
\\
& h^t(\widetilde\rho^t(a,x)) = \rho^t(a,h^t(x)) + m^t(f^t(a), h^t(x)) - m^t(h^t(x), f^t(a)) , 
\\
& h^t(\widetilde m^t(x,y)) =  m^t(h^t(x), h^t(y)) .
\fe

At the first order in $t$,  (\ref{jac}) reduces to the following conditions on $\varphi_1, \rho_1, m_1$:
\ie\label{jacone}
& \varphi_1(a\wedge [b,c]) + [a,\varphi_1(b\wedge c)] + ({\rm cyclic~permutations}) = 0,
\\
& [a, \rho_1(b,x)] + \rho_1(a,[b,x]) - [b,\rho_1(a,x)] - \rho_1(b, [a,x]) = \rho_1([a,b],x) + [\varphi_1(a\wedge b),x],
\\
& [a, m_1(x,y)] + \rho_1(a, xy) = \rho_1(a,x) y + m_1([a,x] ,y) + x\rho_1(a,y)+m_1(x,[a,y]),
\\
& m_1(x,y) z+m_1(xy,z) = xm_1(y,z) + m_1(x,yz).
\fe
The first equation is the cocycle condition on $\varphi_1\in{\rm Hom}(\Lambda^2\cG, N)$, which defines a cohomology class in ${\rm H}^2(\cG, N)$. The equivalence relations (\ref{auto}) on the other hand reduces at first order in $t$ to the following trivial deformations
\ie\label{jacvar}
& \delta \varphi_1(a\wedge b) =  [a, f_1(b)] - [b, f_1(a)] - f_1([a,b]) ,
\\
%ying
%&\delta\rho_1(a,x) = [a,h_1(x)] + [f_1(a), x] -  h_1([a,x]), 
&\delta\rho_1(a,x) = [a,h_1(x)] -  h_1([a,x]) + f_1(a) x - x f_1(a), 
\\
& \delta m_1(x,y) = h_1(x) y + x h_1(y) - h_1(xy) .
\fe

The underlying algebraic structure of the deformation can be understood in terms of the Lie-Hochschild cohomology \cite{Movshev:2009ba} with respect to an $L_\infty$ action of Lie algebras. Let us begin with the Hochschild cochain complex $\widehat {\rm C}^n(N,N)$. There is a natural $\cG$-action, defined on $m\in \widehat{\rm C}^n(N,N)$ as
\ie\label{gdot}
(g\cdot m)(x_1,\cdots, x_n)=[g,m(x_1,\cdots,x_n)]-\sum_{i=1}^nm(x_1,\cdots,[g,x_i],\cdots,x_n).
\fe
There is a different action by ${\cal H}$ on $\widehat{\rm C}^{n+1}(N,N)$,
\ie
(\ell_h\cdot m)(x_1,\cdots, x_{n})=\sum^{n}_{i=0}(-1)^im(x_1,\cdots,x_{i},h,x_{i+1},\cdots,x_{n}).
\fe
It is straightforward to verify that 
\ie\label{LIYYY}
&(h\cdot m)=d_H(\ell_h\cdot m)+\ell_h\cdot d_Hm\equiv\{d_H,\ell_h\}\cdot m,
\\
&\ell_{h_1}\cdot( \ell_{h_2}\cdot m)+\ell_{h_2}\cdot( \ell_{h_1}\cdot m)=0,
\\
&g\cdot (\ell_{h}\cdot m)-\ell_{h}\cdot (g\cdot m)=\ell_{[g,h]}\cdot m,
\\
&g\cdot d_Hm-d_H(g\cdot m)=0,
\fe
where $d_H$ is the Hochschild differential. The action by $\cG$ induces an $L_\infty$-action by $\cG/\cH$ on $\widehat{\rm C}^n(N,N)$ as a differential graded module. Note that for our application ($\cG=L$, $\cH=YM_d$, $\cG/\cH={\bf susy}_d$), the extension $\cH\to \cG\to \cG/\cH$ splits, i.e., there exists a map $ i:\cG/\cH\to \cG$. Given $\{q_\A\}$ a basis of $\cG/\cH$ with Lie bracket $[q_\A,q_\B]=f_{\A\B}^\C q_\C$, where $f^\C_{\A\B}$ are structure constants of ${\cal G} / {\cal H}$, and $i(q_\A)=g_\A \in \cG$, we will define the action of $q_\A$ on $\widehat C(N,N)$ to be the same as that of $g_\A$. Such an action by $q_\A$ does not preserve the Lie algebra structure of $\cG/\cH$, but is rather an $L_\infty$ action, namely
\ie\label{SOMCE}
q_\A\cdot (q_\B\cdot m)-q_\B\cdot (q_\A\cdot m)=f_{\A\B}^\C q_\C\cdot m+\{d_H,\ell_{q_{\A\B}}\}\cdot m,
\fe
for some $q_{\A\B}\in \cH$ (see example below). Now consider the complex $\bigoplus_{p+q=n}\Lambda^p[t^\A]\otimes \widehat{\rm C}^q(N,N)$ equipped with the differential
\ie\label{linf}
d_H-\widehat q+{1\over 2}f^\A_{\B\C}t^\B t^\C {\partial\over\partial t^\A},
\fe
where $\widehat q=\widehat q_\A t^\A+{1\over 2}\widehat q_{\A\B} t^\A t^\B$. The $t^\A$'s are ghost variables dual to $q_\A$, with opposite statistics ($t^\A$ is odd if $q_\A$ is even and vice versa). The hat notation emphasizes the $L_\infty$ action, namely $\widehat q_\A$ acts by $q_\A\,\cdot\,$, and $\widehat q_{\A\B}$ acts as $\ell_{q_{\A\B}}$. One can verify that this differential is indeed nilpotent. The nilpotency condition takes the form of a Maurer-Cartan equation 
\ie\label{MCE}
{1\over 2}f^\A_{\B\C}t^\B t^\C {\partial\over\partial t^\A} \widehat q+\{d_H,\widehat q\}-{1\over 2}\{\widehat q,\widehat q\} = 0.
\fe
%The last equation of \eqref{LIYYY} gives the first order of the Maurer-Cartan equation \eqref{MCE}. The equation \eqref{SOMCE} gives the second order \eqref{MCE}. The third equation of \eqref{LIYYY} gives the third order \eqref{MCE}.

For example, in the application to the IKKT matrix model in zero dimension, $\cG=L$, $\cH=YM$, $\cG/\cH = {\bf susy}_0$, and $i:{\bf susy}_0 \stackrel{\sim}{\longrightarrow} L^1$. The generators $q_\A = D_\A$ are odd, and so the commutators above are to be replaced by the appropriate anti-commutators. We have
\ie
2 D_{(\A}\cdot (D_{\B)}\cdot m) = \Gamma^i_{\A\B}\{d_H,\ell_{D_i}\}\cdot m
\fe
on a Hochschild cochain $m$, and so $q_{\A\B} = \Gamma^i_{\A\B} D_i$.

The purpose of introducing this $L_\infty$ machinery is so that the first order deformations (\ref{jacone}) modulo (\ref{jacvar}) can be rephrased as the degree $n=2$ component of the Lie-Hochschild cohomology
${\rm HH}^n_{L_\infty,\cH}(\cG, \widehat C(N,N))$
defined on the complex
\ie\label{pccc}
\bigoplus_{p+q=n}{\rm Hom}_\cH(\Lambda^p \cG, \widehat C^q(N,N) ) \simeq \bigoplus_{p+q = n} \left( {\Lambda}^p[\eta^I] \otimes \widehat C^q(N,N) \right)_\cH
\fe
with the differential $d_H-\widehat q_I\eta^I+{1\over 2} f^I_{JK} \eta^J \eta^K \partial_{\eta^I}$, where $\eta^I$ are ghost variables dual to the basis $\{ g_I \}$ of $\cG$, $f^I_{JK}$ are the structure constants of $\cG$, and $\widehat q_I$ acts by $g_I \, \cdot$ as defined in (\ref{gdot}). A subset of these, $\{g_a \}$, generate the ideal $\cH$. The subscript $\cH$ indicates that the cochain $f$ in (\ref{pccc}) are subject to the $\cH$-invariance condition\footnote{The $\cH$-invariance condition is preserved by the differential,
\ie
\{(d_H-\widehat q_I\eta^I+{1\over 2} f^I_{JK} \eta^J \eta^K \partial_{\eta^I}),\left(x^a \partial_{\eta^a}+\ell_h\right)\}=-f^b_{Ia}\eta^Ix^a\left(\partial_{\eta^b}+\ell_{g_b}\right).
\fe}
\ie
\left(x^a {\partial\over \partial\eta^a}+\ell_h\right) f = 0,
\fe
for all $h=x^a g_a\in\cH$, which ensures the compatibility condition (\ref{compc}) between various deformation maps.

On the other hand, the Lie algebra cohomology ${\rm H}^n(\cG, N)$, defined in terms of Chevalley-Eilenberg cochain complex, can be reformulated via the Hochschild-Serre spectral sequence as the Lie-Hochschild cohomology with respect to the $L_\infty$ action of $\cG/\cH$, namely ${\rm HH}^n_{L_\infty}(\cG/\cH, \widehat C(N,N))$ defined from the complex
\ie\label{qccc}
\bigoplus_{p+q=n}{\rm Hom}(\Lambda^p (\cG/\cH), \widehat C^q(N,N) ) \simeq \bigoplus_{p+q = n} {\Lambda}^p[t^\A] \otimes \widehat C^q(N,N)
\fe
with the differential (\ref{linf}). The two cochain complexes  (\ref{pccc}) and (\ref{qccc}) are isomorphic,\footnote{
The differential on the complex (2.36) when restricting to $\eta^a=0$ becomes the differential on the complex (2.39).
}
which implies that the inequivalent triples $(\varphi_1, \rho_1, m_1)$ are indeed classified by $[\varphi_1]\in {\rm H}^2(\cG, N)$ alone. The obstruction class to second order deformation is given in terms of the Gerstenhaber bracket of $(\varphi_1,\rho_1,m_1)$ with itself, which lies in ${\rm H}^3(\cG, N)$. In our case, $\cG=L$, $N=U(YM_d)$, we will reformulate this construction explicitly in terms of pure spinor variables and discuss some examples in section \ref{higherorder}.

\section{A classification of infinitesimal deformations}

Now we summarize the results of our classification of the F-term\footnote{
Our classification of F-term deformations is based on several assumptions used in computing the hypercohomology in Appendix~\ref{Hypercohomology}.  Additional F-term deformaitons may exist if some of the assumptions fail.  Our classification of exceptional D-terms is complete.
}
and exceptional D-term deformations. The details of the computation that led to this classification are explained in Appendix~\ref{Classification}. First we describe the deformations that are invariant under both Lorentz and R-symmetries, in every spacetime dimension from 0 to 10. Then we describe the (still fully supersymmetric) deformations that are Lorentz invariant but not R-symmetry singlets, and the ones that are R-symmetry invariant but not Lorentz singlets.

\subsection{$SO(d)\times SO(10-d)$ invariant deformations}
%In this subsection, we list all the deformations that are invariant under the $SO(d)$ Lorentz and $SO(10-d)$ R-symmetry.

\subsubsection{F-term deformations}
The only F-term deformation that preserves the full $SO(d)\times SO(10-d)$ symmetry corresponds to a cohomology class in ${\rm H}^2(L,{\rm Sym}^3(YM_d))_8$. In the complex ${\rm Sym} (YM_d)\otimes\cS$ of \eqref{psf}, this class can be represented by
\ie
(\lambda \Gamma^m\chi)\circ (\lambda\Gamma^n\chi)\circ F_{mn},
\fe
where $\circ$ denotes the symmetric product. This is the well-known Born-Infeld deformation.

\subsubsection{Exceptional D-term deformations}\label{EDtermD}

There are two exceptional D-term deformations that preserve $SO(10)$ in 10 dimensions, and one exceptional D-term deformation that preserves $SO(8)\times SO(2)$ in 8 dimensions.

The first exceptional D-term deformation in 10 spacetime dimensions corresponds to a class in ${\rm H}_1(L,{\rm Sym}^1(YM_{10}))_4$, represented by the cycle
\ie\label{ED10a}
 D_\A\otimes \chi^\A.
\fe
\eqref{ED10a} maps to a nontrivial class in ${\rm H}_1(L,{\rm Sym}^1(YM))$ under the map induced by the inclusion $YM_{10}\subset YM$, and it can be pulled back to a class
\ie
%ying
%D_\A\otimes \chi^\A = {4\over 5}B_L
 {4\over 5} \la D_m\circ D_m \ra
\fe
in ${\rm H}_0(L,{\rm Sym}^2(YM))_4$ under $B_{YM}$.
By the commutativity of the diagram \eqref{commdiag} at $d=0$, the Lagrangian density of this deformation is given by
\ie
A_0 \tr(D_m\circ D_m) = \epsilon^{\A_1\cdots\A_{16}} D_{\A_1}\cdots D_{\A_{16}} \tr(D_m\circ D_m) .
\fe
In the language of the component field Lagrangian, this deformation corresponds to a dimension 10 operator. Interestingly, its reduction to lower spacetime dimensions (in which case it becomes an ordinary D-term) appears to be the counterterm responsible for the 2-loop divergence in 7-dimensional MSYM, the 3-loop divergence of 6-dimensional MSYM, and the 6-loop divergence of the 5-dimensional MSYM.

The second exceptional D-term deformations in 10 dimensions corresponds to a class in ${\rm H}_1(L,{\rm Sym}^3(YM_{10}))_{12}$, represented by the following cycle
\ie\label{ED10b}
14D_\A\otimes \chi^\A\circ F_{mn}\circ F^{mn}-D_\A\otimes(\Gamma^{mnpq}\chi)^\A\circ F_{mn}\circ F_{pq},
\fe
which is mapped to a nontrivial class in ${\rm H}_1(L,{\rm Sym}^3(YM ))_{12}$ and can be further pulled back to a class
\ie
\la 2D_p\circ D^p\circ F_{mn}\circ F^{mn}-3D_p\circ F_{mn}\circ \chi^\A\circ (\Gamma^{mnp}\chi)_\A\ra
\fe
 in ${\rm H}_0(L,{\rm Sym}^4(YM)_{12}$. The Lagrangian density is given by
\ie
\epsilon^{\A_1\cdots\A_{16}} D_{\A_1}\cdots D_{\A_{16}}\tr(2D_p\circ D^p\circ F_{mn}\circ F^{mn}-3D_p\circ F_{mn}\circ \chi^\A\circ (\Gamma^{mnp}\chi)_\A).
\fe
In the component field Lagrangian this corresponds to a dimension 14 operator.

The exceptional D-term deformation in 8 spacetime dimensions corresponds to a class in ${\rm H}_1(L,{\rm Sym}^2(YM_8))_{8}$, represented by the cycle
\ie\label{ED8}
14D_\A \otimes \chi^\A \circ F_{9,10} - D_\A \otimes (\Gamma_{9,10,\m\n}\chi)^\A \circ F_{\m\n}.
\fe
This class maps to a nontrivial class in ${\rm H}_1(L,{\rm Sym}^2(YM))_{8}$ under the the map induced by the inclusion $YM_8\subset YM$, and it can further be pulled back to a class 
\ie
\vev{D_p\circ \chi^\A\circ (\Gamma_{9,10,p}\chi)_\A}
\fe
in ${\rm H}_0(L,{\rm Sym}^3(YM))_8$. The Lagrangian density of this deformation is then given by
\ie
\epsilon^{\A_1\cdots\A_{16}} D_{\A_1}\cdots D_{\A_{16}}\tr(D_\m\circ \chi^\A\circ (\Gamma_{9,10,\m}\chi)_\A).
\fe 
In the component field Lagrangian this corresponds to a dimension 12 operator. %Interestingly, this operator appears to be the counterterm responsible for the 2-loop divergence in 8-dimensional MSYM, at the perturbative quantum level.

\subsection{Lorentz invariant deformations}

In this subsection, we list all the fully supersymmetric single trace deformations of $d$-dimensional MSYM that preserve the $SO(d)$ Lorentz symmetry, but break the $SO(10-d)$ R-symmetry.

\subsubsection{F-term deformations}

There is an F-term deformation in each symmetric $k$-tensor representation of $SO(10-d)$ R-symmetry. It corresponds to a class in ${\rm H}^2(L,{\rm Sym}^{k+3}(YM_d))_{2k+8}$. In the complex ${\rm Sym}(YM_d)\otimes\cS$ in \eqref{psf}, this class can be represented by the cocycle
\ie
 (\lambda \Gamma^m \chi) \circ (\lambda \Gamma^n \chi) \circ (\chi \circ \Gamma_{mn(i_1} \chi) \circ D_{i_2} \circ \dotsb \circ D_{i_k)}, \quad k \geq 1.
\fe
In the component field language, they correspond to Lagrangian deformations by 8 supercharges acting on a half BPS operator. In this sense they can be thought of as half superspace integrals, just like the Born-Infeld deformation.

There is an extra F-term deformation in 0-dimensional MSYM (i.e. IKKT matrix model) in the self-dual 5-form representation of the $SO(10)$ R-symmetry. It corresponds to a class in ${\rm H}^2(L,{\rm Sym}^{2}(YM))_{2}$. In the complex ${\rm Sym}(YM)\otimes\cS$ in \eqref{psf}, this class can be represented by the cocycle
\ie
(\lambda \Gamma_{mnpqr} \lambda) D_s\circ D_s - 10 D_{[m} \circ (\lambda \Gamma_{npqr]s} \lambda) D_s.
\fe
In the component field Lagrangian, it corresponds to 4 supercharges acting on a half BPS operator, of the form
\ie
& (\Gamma_{[\underline{ab}}{}^m)^{\A\B}(\Gamma_{\underline{cd}}{}^n)^{\C\D} D_\A D_\B D_\C D_\D ~ {\rm tr} (\Phi_{\underline{e}]}\circ \Phi_m \circ \Phi_n)'
\\
& \sim {\rm tr} \left(\Phi_{[a}\circ [\Phi_b,\Phi_c]\circ[\Phi_d,\Phi_{e]}]+\cdots\right).
\fe
The prime in the first line indicates that the traces on the vector indices of the three $\Phi$'s are removed. This deformation arises in the world volume theory of multi-D-instantons in type IIB string theory on $AdS_5\times S^5$, with the $AdS_5\times S^5$ viewed as a deformation from flat spacetime.

\subsubsection{Exceptional D-term deformations}
There is no Lorentz symmetry preserving, but R-symmetry breaking exceptional D-term deformation.

\subsection{R-symmetry invariant deformations}

In this subsection, we list all the fully supersymmetric single trace deformations that preserve $SO(10-d)$ R-symmetry, while breaking $SO(d)$ Lorentz symmetry.

\subsubsection{F-term deformations}

There is a class of F-term deformations in every spacetime dimension $d$, that transforms in the anti-symmetric 2-form representation of the $SO(d)$ Lorentz symmetry. It corresponds to a class in ${\rm H}^2(L,{\rm Sym}^{2}(YM_d))_{4}$. In the complex ${\rm Sym}(YM_d)\otimes\cS$ in \eqref{psf}, this class can be represented by the cocycle
\ie
(\lambda\Gamma_i\chi)\circ(\lambda\Gamma_j\chi).
\fe
This is the usual noncommutative deformation of MSYM theories.

\subsubsection{Exceptional D-term deformations}

R-symmetry invariant exceptional D-term deformations exist in spacetime dimension 8, 9 and 10. The one in 8 spacetime dimensions is an $SO(8)\times SO(2)$ singlet, and has been already discussed in Section \ref{EDtermD}. 

In 9 spacetime dimensions, there is an exceptional D-term deformation in the vector representation of $SO(9)$. It corresponds to a class in ${\rm H}_1(L,{\rm Sym}^2(YM_9))_8$, and can be represented by the cycle
\ie
 14D_\A \otimes \chi^\A \circ F_{i,10} - D_\A \otimes (\Gamma_{i,10,pq}\chi)^\A \circ F_{pq}.
\fe

In 10 spacetime dimensions, there is an exceptional D-term deformation in the anti-symmetric 2-form representation, and an exceptional D-term deformation in the self-dual 5-form representation. The exceptional D-term deformation in the 2-form representation corresponds to a class in ${\rm H}_1(L,{\rm Sym}^2(YM_{10}))_8$, and can be represented by the cycle
\ie
 14D_\A \otimes \chi^\A \circ F_{mn} - D_\A \otimes (\Gamma_{mnpq}\chi)^\A \circ F_{pq}.
\fe
The Lorentz 5-form deformation corresponds to a class in ${\rm H}_1(L,{\rm Sym}^4(YM_{10}))_{14}$.

\section{Higher order deformations}
\label{higherorder}

In the on-shell formulation, it is a very nontrivial problem to extend the infinitesimal supersymmetric deformations beyond the first order. A priori, there can be obstructions that correspond to cohomology classes in ${\rm H}^3(L,U(YM_d))$, as we have seen in section \ref{formald}. While such obstructions can in principle be computed as Gerstenhaber brackets, and the higher order deformation can be determined when the obstruction class vanishes, in practice a direct computation is very difficult, partly due to the complicated form of the inverse Cartan-Eilenberg map.

One way to compute the obstruction classes and higher order deformations is to enlarge the $L$-module $U(YM_d)$, or the complex $U(YM_d)\otimes {\cal S}$, in such way that the cohomology class representing the infinitesimal deformation is trivialized. One can then absorb the deformation by a redefinition of the generators of $L$ in the enlarged module. This allows for a construction of all order deformations in the enlarged module. One then tries to show that these higher order deformations are cohomologically equivalent to ones that lie in the original complex $U(YM_d)\otimes{\cal S}$.

In this section we describe some limited progress along this line. The structure we uncover here seems closely related to the non-minimal extension of the pure spinor formalism. Ultimately, the best way to determine the higher order deformation is based on the off-shell formulation (where the non-minimal pure spinor superspace is employed), which will be the subject of a companion paper.

\subsection{Obstruction classes and the non-minimal pure spinor formalism}

Let us begin with the deformed product on the generators of $U(L)$,
\ie\label{dfp}
\lambda^\A \lambda^\B D_\A \star D_\B = \epsilon {\cal O}_\lambda+{\cal O}(\epsilon^2),
\fe
where ${\cal O}_\lambda \equiv \lambda^\A \lambda^\B {\cal O}_{\A\B}$, ${\cal O}_{\A\B}\in U(YM_d)$. Associativity at first order in $\epsilon$ demands that
${\cal O}_\lambda$ obeys the cocycle condition on $U(YM_d)\otimes {\cal S}$,
\ie{}
\left[ Q, {\cal O}_\lambda \right] = 0.
\fe
The question is to extend the deformation to higher orders in $\epsilon$ while maintaining the associativity of $\star$, by adding operators on the RHS that take value in $U(YM_d)$.

In some simple cases, such as the noncommutative deformation, ${\cal O}_\lambda$ will become exact once we extend the module from $N=U(YM_d)$ to $U(YM)$, or to $U(L)$. Generally this is not enough: the cocycle ${\cal O}_\lambda$ may not be exact in $U(L)\otimes{\cal S}$ either, as is the case for the Born-Infeld deformation. The idea is to further enlarge the module $N\otimes {\cal S}\subset U(L)\otimes{\cal S}$ to some ${\cal N}$ so as to trivialize $Q$-cohomology, so that ${\cal O}_\lambda$ becomes an exact element in ${\cal N}$. 
This can be achieved by introducing non-minimal pure spinor variables $\olam_\A$, and taking 
\ie
{\cal N} = U(L)\otimes {\cal S}_{\lambda,\olam},
\fe
where ${\cal S}_{\lambda,\olam}$ is the ring of polynomials in the pure spinors $\lambda^\A$, $\olam_\A$, as well as $(\lambda\olam)^{-1}$.
We will see later that for the Born-Infeld deformation, ${\cal O}_\lambda$ is indeed exact in ${\cal N}$. For now let us assume this is the case, and write
\ie
{\cal O}_\lambda = \{ Q, R\}
\fe
for some $R = \lambda^\A R_\A$ in ${\cal N}$. There is an ambiguity of shifting $R$ by a $Q$-exact element,
\ie
\delta R = [Q, \Omega],~~~~\Omega\in{\cal N}.
\fe
%We will make use of this shift momentarily.
Now consider a redefinition of generators,
\ie\label{dtt}
\widetilde D_\A = D_\A - \epsilon R_\A,
\fe
so that the $\widetilde D_\A$'s under the deformed $\star$ product obey the same quadratic relation as $D_\A$'s under the undeformed product, up to ${\cal O}(\epsilon^2)$ corrections. Namely, it follows from (\ref{dfp}) and (\ref{dtt}) that 
\ie
\lambda^\A \lambda^\B \widetilde D_\A \star \widetilde D_\B = {\cal O}(\epsilon^2).
\fe
This suggests that we construct the higher order deformation of the $\star$ product by demanding that $\widetilde D_\A$'s under $\star$ product obey {exactly} the same relations as $D_\A$'s did under the original undeformed relation of $U(L)$. Namely, we insist on
\ie\label{extd}
\lambda^\A \lambda^\B \widetilde D_\A \star \widetilde D_\B = 0,
\fe
and that $D_\A$ is related by 
\ie\label{drla}
D_\A = \widetilde D_\A + \epsilon \widetilde R_\A .
\fe
Now we will view $\widetilde R_\A$ as an expression built out of $\star$ products of $\widetilde D_\B$. By virtue of (\ref{extd}), the level 2 and higher elements in the Lie algebra generated by $\widetilde D_\A$ under $\star$-commutator can be {\it exactly identified} as the Lie algebra $YM$. By doing so, we have then completely specified $\star$ as a deformed product on $U(L)\otimes {\cal S}_{\lambda,\olam}$. Note that $Q\star Q$ is generally not an element of $U(YM)\otimes {\cal S}$, and we haven't yet found a true deformation of $U(L)$.
What we have is
\ie
\lambda^\A \lambda^\B D_\A \star D_\B = \epsilon \widetilde {\cal O}_\lambda + \epsilon^2 \widetilde R\star \widetilde R,
\fe
where $\widetilde {\cal O}_\lambda = \{\lambda^\A\widetilde D_\A, \widetilde R\}_\star$. A key point is that $\widetilde {\cal O}_\lambda$ is an element of $U(YM)\otimes {\cal S}_2$ (the subscript stands for the degree in $\lambda$), and is independent of the non-minimal variable $\olam$. 
Since $\widetilde R=\lambda^\A \widetilde R_\A$ is built out of $\widetilde D_\A$, $\widetilde R\star\widetilde R$ can be computed by applying the relations on $\widetilde D_\A$ under $\star$ that take the same form as the relations as obeyed by $D_\A$ under the original undeformed product in $U(L)$. The question is whether $\widetilde R \star \widetilde R$ is also in $U(YM)\otimes {\cal S}_2$. If this is true, then we already have an all-order deformation of the superfield equation of motion, as desired. But this won't be the case in general. If the second order deformation is indeed unobstructed, we can only expect that, after some appropriate shift $\delta R = [Q,\Omega]$, we can find an $R$ such that
\ie
\left[ \widetilde {\cal O}_\lambda, \widetilde R \right]_\star =\left[ \lambda^\A \widetilde D_\A, \widetilde R\star \widetilde R \right]_\star =  \left[ \lambda^\A \widetilde D_\A, \widetilde {\cal O}^{(2)}_\lambda \right]_\star,
\fe
where $\widetilde {\cal O}^{(2)}_\lambda \in U(YM)\otimes {\cal S}_2$. In other words, $\widetilde R\star\widetilde R$ is cohomologous to $\widetilde {\cal O}^{(2)}$ which is independent of $\olam$.
If the cohomology of $\big[\lambda^\A \widetilde D_\A,\,\cdot\, \big]_\star$ is trivial, then we would be able to find $R^{(2)}= \lambda^\A \widetilde R^{(2)}_\A$ in ${\cal N}$ such that
\ie
\widetilde R\star \widetilde R = \widetilde {\cal O}^{(2)}_\lambda - \left\{ \lambda^\A \widetilde D_\A , \widetilde R^{(2)} \right\}_\star.
\fe 
This leads to
\ie
\lambda^\A \lambda^\B (\widetilde D_\A + \epsilon \widetilde R_\A + \epsilon^2 \widetilde R^{(2)}_\A) \star (\widetilde D_\B + \epsilon \widetilde R_\B+ \epsilon^2 \widetilde R^{(2)}_\B) = \epsilon \widetilde {\cal O}_\lambda + \epsilon^2 \widetilde {\cal O}^{(2)}_\lambda + {\cal O}(\epsilon^3).
\fe
We are then instructed to correct the relation (\ref{drla}) by adding an order $\epsilon^2$ term,
\ie
D_\A = \widetilde D_\A + \epsilon\widetilde R_\A + \epsilon^2 \widetilde R^{(2)}_\A,
\fe
while still insisting on $\widetilde D_\A$ themselves obey the same relations under $\star$. This amounts to correcting the $\star$ product of $D_\A$ at order $\epsilon^2$, to
\ie
\lambda^\A \lambda^\B D_\A\star D_\B = \epsilon \widetilde {\cal O}_\lambda + \epsilon^2 \widetilde {\cal O}^{(2)}_\lambda + \epsilon^3\left\{ R, R^{(2)} \right\}_\star +{\cal O}(\epsilon^4).
\fe
We then carry on the same procedure, and ask if we can find a $\lambda^\A\widetilde D_\A$-exact shift of $\widetilde R^{(2)}$ such that
\ie
\left[ \lambda^\A \widetilde D_\A, \{\widetilde R, \widetilde R^{(2)} \}_\star \right]_\star = \left[ \lambda^\A \widetilde D_\A, \widetilde {\cal O}^{(3)}_\lambda \right]_\star,
\fe
for some $\widetilde {\cal O}^{(3)}_\lambda\in U(YM)\otimes {\cal S}_2$.
If so, we seek $\widetilde R^{(3)} = \lambda^\A \widetilde R^{(3)}_\A \in {\cal N}$, that obeys
\ie
\{\widetilde R_\lambda, \widetilde R^{(2)}_\lambda\}_\star = \widetilde {\cal O}^{(3)}_\lambda - \left\{ \lambda^\A \widetilde D_\A , \widetilde R^{(3)} \right\}_\star,
\fe
and so on and so forth.

For the computation of these higher order deformations, we might as well drop all $\sim$ scripts and at the same time replace $\star$ by the original undeformed product of $U(L)$ or its non-minimal extension ${\cal N}=U(L)\otimes {\cal S}_{\lambda,\olam}$. 

To summarize, in order to show that there is no obstruction, and to construct the next order deformations, we need to first write ${\cal O}_\lambda$ in the form $\{Q, R\}$ for some $R = \lambda^\A R_\A\in {\cal N}$, then find an ${\cal O}^{(2)}_\lambda \in U(YM)\otimes {\cal S}_2$, namely one that is independent of $\olam$, such that
\ie{}
\left[ Q, (R + [Q, \Omega])^2 \right] = [{\cal O}_\lambda, R + [Q, \Omega]] = \left[ Q, {\cal O}^{(2)}_\lambda \right].
\fe
The freedom of shifting $R$ by $[Q, \Omega]$ amounts to shifting ${\cal O}_\lambda^{(2)}$ by
\ie
{\cal O}^{(2)}_\lambda\to {\cal O}^{(2)}_\lambda + [{\cal O}_\lambda, \Omega].
\fe
Next, we try to find $R^{(2)} = \lambda^\A R^{(2)}_\A$ in ${\cal N}$, such that
\ie
R\cdot R = {\cal O}^{(2)}_\lambda + \left\{ Q, R^{(2)} \right\},
\fe
and so forth. We now give a few examples of such computations.

\subsection{Examples of second order deformations}

\subsubsection{Noncommutative deformation}

As already seen, the noncommutative deformation of MSYM at the first order is represented by the cocycle in ${\rm Sym}(YM_d)\otimes {\cal S}$,
\ie
{\cal O}_\lambda^{NC} = \omega^{mn} (\lambda\Gamma_m \chi) \circ (\lambda\Gamma_n \chi).
\fe
Here $m,n$ are along the $d$ spacetime directions.
Now if we regard ${\cal O}^{NC}_\lambda$ as an element of ${\rm Sym}(YM)\otimes {\cal S}$, it becomes $Q$-exact, with ${\cal O}^{NC}_\lambda = \{Q, R\}$, and
\ie
R = \omega^{mn} D_m \circ (\lambda\Gamma_n \chi).
\fe
Indeed while ${\cal O}_{\A\B}^{NC}$ is an element of ${\rm Sym}(YM_d)$, $R_\A$ lies in ${\rm Sym}(YM)$ due to the $D_m$ factor in the symmetrized product.

To compute $R^2$, and express the result in terms of symmetrized products, we can make use of Baker-Campbell-Hausdorff formula, and in particular
\ie
& \left.\exp(\ln (e^X e^Y)) - \exp(\ln (e^Y e^X)) \right|_{X^2 Y^2}
\\
& = X\circ Y\circ [X,Y] - {1\over 24} [X,[Y,[X,Y]]] - {1\over 24}[Y,[X,[X,Y]]].
\fe
Any factor that involves a commutator appearing in $R^2$ already lies in $YM_d$. All we need to worry about is the term $X\circ Y\circ [X, Y]$ which is a priori an element of ${\rm Sym}^3(YM)$ but not ${\rm Sym}^3(YM_d)$. A simple computation gives
\ie
& R^2 = \omega^{mn} \omega^{pq} \bigg[ D_m \circ D_p \circ \{ \lambda\Gamma_n \chi, \lambda\Gamma_q\chi\} + 2 D_m\circ (\lambda\Gamma_p\chi)\circ (\lambda\Gamma_n D_q \chi) +(\lambda\Gamma_m\chi)\circ(\lambda\Gamma_p\chi)\circ[D_n,D_q]\bigg]
\\
&~~~~~~~ +{\cal O}^{(2)}_\lambda 
\\
&~~~=Q\bigg[\omega^{mn} \omega^{pq} (D_m\circ D_p\circ[D_n,\lambda\Gamma_q\chi]+D_m\circ (\lambda\Gamma_p\chi)\circ[D_n,D_q])\bigg]+{\cal O}^{(2)}_\lambda,
\fe
where ${\cal O}_\lambda^{(2)}\in {\rm Sym}^3(YM_d)$.
From this we can also read off $\widetilde R^{(2)}_\lambda$,
\ie
\widetilde R^{(2)}_\lambda=\omega^{mn} \omega^{pq} (D_m\circ D_p\circ[D_n,\lambda\Gamma_q\chi]+D_m\circ (\lambda\Gamma_p\chi)\circ[D_n,D_q]).
\fe

\subsubsection{5-form deformation}

As stated earlier in our classification, the 0-dimension MSYM has an F-term deformation that transforms in the self-dual 5-form representation of the $SO(10)$ R-symmetry, represented by the cocycle in ${\rm Sym}(YM)\otimes {\cal S}$:
\ie
\cO^{SD} &= {1\over 16\cdot 5!}\omega^{\A\B} (\Gamma^{abcde})_{\A\B}\left[ (\lambda \Gamma_{abcde} \lambda) D^2 - 10 D_{[a} \circ (\lambda \Gamma_{bcde]f} \lambda) D^f \right]
\\
&=-\omega^{\A\B} QD_\A \circ QD_\B,
\fe
where $\omega^{(\A\B)}$ transforms in the representation $[00002]$ of $Spin(10)$.
While ${\cal O}^{abcde}$ represents a nontrivial cohomology class in ${\rm H}^2({\rm Sym}(YM)\otimes {\cal S})$, it becomes becomes $Q$-exact in ${\rm Sym}(L)\otimes {\cal S}$. We have ${\cal O}^{SD} = \{Q, R\}$, with
\ie
R= \omega^{\A\B} D_\A\circ QD_\B.
\fe
We can then compute
\ie
R^2&=-\omega^{\A\B} \omega^{\C\D} \bigg[ D_\A \circ D_\C \circ [QD_\B, QD_\D] - 2 D_\A\circ QD_\C\circ [QD_\B,D_\D] +QD_\A\circ QD_\C\circ\{D_\B,D_\D\}\bigg]
\\
&~~~ +{\cal O}^{(2)}_\lambda 
\\
&=-Q\bigg[\omega^{\A\B} \omega^{\C\D} \big(D_\A\circ D_\C\circ[D_\B,QD_\D]+D_\A\circ QD_\C\circ\{D_\B,D_\D\}\big)\bigg]+{\cal O}^{(2)}_\lambda,
\fe
and find
\ie
\widetilde R^{(2)}_\lambda = -\omega^{\A\B} \omega^{\C\D} \Big(D_\A\circ D_\C\circ[D_\B,QD_\D]+ D_\A\circ QD_\C\circ\{D_\B,D_\D\}\Big).
\fe

\subsubsection{Born-Infeld deformation}

The first order Born-Infeld deformation is represented by the cocycle 
\ie
{\cal O}_\lambda = (\lambda \Gamma^m \chi)\circ (\lambda\Gamma^n \chi) \circ F_{mn} = QD^m\circ QD^n\circ F_{mn}.
\fe
In order to trivialize the cohomology of ${\cal O}_\lambda$, extending ${\rm Sym}(YM)$ to ${\rm Sym}(L)$ is not enough. We need to consider the module ${\cal N}={\rm Sym}(L)\otimes{\cal S}_{\lambda,\olam}$, by allowing dependence on the non-minimal pure spinor variable $\olam_\A$, as well as $(\lambda\olam)^{-1}$. Using pure spinor identities, one can verify that
\ie
R = {1\over 2} (\lambda\olam)^{-1} (\lambda \Gamma^m \chi)\circ (\lambda\Gamma^n \chi) \circ (\olam\Gamma_{mn} \chi)
\fe
obeys $\{Q, R\} = {\cal O}_\lambda$.
%\ie
%\{Q, R\} &= -{1\over 8} (\lambda\olam)^{-1} (\lambda \Gamma^m \chi)\circ (\lambda\Gamma^n \chi) \circ (\olam\Gamma_{mn}\Gamma^{pq}\lambda) F_{pq}
%\\
%&= -{1\over 8} (\lambda\olam)^{-1} (\lambda \Gamma^m \chi)\circ (\lambda\Gamma^n \chi) \circ (\olam[\Gamma_{mn}, \Gamma^{pq}]\lambda) F_{pq}
%\\
%&= - (\lambda\olam)^{-1} (\lambda \Gamma^m \chi)\circ (\lambda\Gamma^n \chi) \circ (\olam\Gamma_{mq}\lambda) F_n{}^q
%\\
%&= (\lambda \Gamma^m \chi)\circ (\lambda\Gamma^n \chi) \circ F_{mn}.
%\fe
Keep in mind that, in order to find the second order deformation, we may need to further shift
\ie
R \to R + [Q, \Omega],
\fe
for some $\Omega$ of homogeneous degree zero in $\lambda$ and $\olam$.

In principle, we would like to compute $R\cdot R$ (keep in mind that $R$ is odd and $R^2$ is a nontrivial anti-commutator),
and express the result in the form of a symmetrized product.
%\ie
%R = {1\over 2}(\lambda\olam)^{-1} (\lambda\Gamma^m)_\A (\lambda\Gamma^n)_\B (\olam\Gamma_{mn})_\C \chi^{[\A}\circ \chi^\B \circ \chi^{\C]}.
%\fe
Since $R$ is the symmetrized product of three $\chi$'s, %all we need to compute is
%\ie
%\left. \exp\left\{ \ln\left[\exp(\epsilon_\A \chi^\A) \exp(\eta_\B \chi^\B)\right] \right\} -  \exp\left\{ \ln\left[\exp(\eta_\B \chi^\B) \exp(\epsilon_\A \chi^\A)\right] \right\} \right|_{\epsilon^3 \eta^3}
%\fe
%where $\epsilon_\A$ and $\eta_\A$ are Grassmannian parameters. The $\ln(e^X e^Y)$ here will be expressed in terms of nested commutators entirely using BCH formula.
%
%Using mathematica (see BCH.nb), we obtain
we can apply the following special case of Baker-Campbell-Hausdorff formula,
\ie
& \left.\exp(\ln (e^X e^Y)) - \exp(\ln (e^Y e^X)) \right|_{X^3 Y^3}
\\
& = {1\over 4} X\circ X\circ Y\circ Y\circ [X,Y] + {1\over 24} [X,Y]\circ [X,Y]\circ [X,Y] + {1\over 12} X\circ [X,Y]\circ [Y,[Y,X]] 
\\
&~~~ + {1\over 12} Y\circ [X,Y]\circ [X,[X,Y]] + {1\over 24} X\circ Y \circ \big( [X,[Y,[Y,X]]] - [Y,[X,[X,Y]]]  \big)
\\
&~~~ - {1\over 180} [X,[Y,[X,[Y,[Y,X]]]]] + {1\over 180} [Y,[X,[Y,[X,[X,Y]]]]] 
\\
&~~~ + {1\over 720} [X,[X,[Y,[Y,[Y,X]]]]] - {1\over 720} [Y,[Y,[X,[X,[X,Y]]]]]
\\
&~~~ - {1\over 720} [X,[Y,[Y,[X,[X,Y]]]]] + {1\over 720} [Y,[X,[X,[Y,[Y,X]]]]].
\fe
%Let us write $R = {1\over 2}(\lambda\olam)^{-1} t_{\A\B\C} \chi^\A \circ\chi^\B\circ \chi^\C$, where $t_{\A\B\C}$ is the anti-symmetric tensor
%\ie
%t_{\A\B\C} =  (\lambda\Gamma^m)_{[\A} (\lambda\Gamma^n)_\B (\olam\Gamma_{mn})_{\C]} .
%\fe
In the end, we would like to write $R^2$ in the form
\ie
R\cdot R = {\cal O}^{(2)}_\lambda + \{ Q, R^{(2)}\} + [{\cal O}_\lambda, \Omega],
\fe
where ${\cal O}^{(2)}_\lambda \in U(YM)\otimes {\cal S}_2$ is independent of $\olam$, and $\Omega$, $R^{(2)}$ are elements of ${\cal N}$ that depend $\olam$. While this should be possible, we have not carried out this computation explicitly. In the on-shell approach considered in this paper, proving the absence of obstruction and finding higher order deformations is generally quite difficult. This question is best addressed in the off-shell formulation, which we consider in the next paper.

\section{Discussion}

All of the F-term deformations listed in our classification are equivalent to Lagrangian deformations by a supersymmetry descendant of a half BPS operator in MSYM, though not necessarily a half superspace integral ($Q^8$-descendant). Presumably the same classification result can also be obtained by directly inspecting the operator spectrum. Our approach, following Movshev and Schwarz, does have the advantage of making all supersymmetries manifest, and allows for writing down the full supersymmetric completion of these terms easily in the superfield formalism.

Potentially, an immediate application of our classification is to the study of SYM theories using supersymmetric localization. It is believed that the six-dimensional $A_{N-1}$ $(2,0)$ superconformal field theory \cite{Aganagic:1997zq, Aharony:1997an}, when compactified on the circle, is a UV completion of the five-dimensional MSYM theory \cite{Henningson:2004ix, Douglas:2010iu}. In some particular renormalization scheme, this 5D theory should be described by the MSYM Lagrangian together with infinitely many higher derivative operators. While it is attempting to conjecture that this higher derivative operators are somehow absent, this possibility appears to be ruled out since the 5D MSYM is after all not perturbatively UV finite \cite{Bern:2012di}. It is then an interesting question what these higher derivative terms are exactly. As we have seen, the only single trace deformations that are invariant under Lorentz and R-symmetry are the Born-Infeld deformation and D-terms. The Born-Infeld deformation is the only one that could affect the computation of e.g. the supersymmetric $S^5$ partition function, which computes the superconformal index of the 6D theory. We are not aware of an argument that rules out the presence of this term in the 5D theory, though it is likely that it is in fact absent.

Another nontrivial example of a UV completion of MSYM occurs in six-dimensions, namely the $A_{N-1}$ $(1,1)$ little string theory \cite{Witten:1995zh, Seiberg:1997zk}, whose low energy limit is the 6D MSYM. In this case, a successful matching of the $F^4$ term in the Coulomb branch effective action \cite{Aharony:2003vk, Aharony:2004xn}, between that of (the undeformed) 6D MSYM, and of the double scaled little string theory \cite{Giveon:1999px, Giveon:1999tq} indicates that the ${\rm Tr} F^4$ term is absent in derivative expansion of the 6D non-Abelian gauge theory in question (at the origin of the Coulomb branch). The possible higher derivative terms must all be D-terms, and it would be interesting to determine them, say by comparing with the double scaled little string theory (though the latter is really approaching the problem from large distances on the Coulomb branch).

The algebraic approach adopted in this paper, in principle, also formulates the problem of finding higher order deformations in a systematic way. Unfortunately, in practice the latter still appears to be a very difficult problem in the deformation theory of associative algebras. In some simple cases, such as the noncommutative deformation and the 5-form deformation in zero dimension, the second order deformation can be found by explicit computation. In some other cases, such as the Born-Infeld deformation, it is possible to prove by inspecting elements of the obstruction cohomology group ${\rm H}^3(L, U(YM_d))$ that there are no candidate obstruction class of the appropriate degree, and thus the deformation can be extended to second order (or $\A'^4$ in the language of open string effective action). This would not be the case for the higher order obstruction classes however, and a direct computation of the potential obstruction class is quite hard, partly due to the complicated explicit form of isomorphism ${\rm H}^*(U(YM_d) \otimes {\cal S})\to {\rm H}^*(L, U(YM_d))$ at th level of cochains. The off-shell formulation of the problem, which will be discussed in our next paper, offers a solution to this problem. We will see there that the (non-Abelian) Born-Infeld deformation can be extended to all orders, based on a BV action written in the non-minimal pure spinor superspace. In principle, the on-shell equation in the superfields can be recovered from it, by eliminating the auxiliary fields.

In this paper we have limited ourselves to single trace deformations. For finite rank gauge groups, or for Abelian theories, the algebraic approach is still possible, if one replaces $YM_d$ by its quotient by an ideal generated by relations among products of finite size matrices. The complex $U(YM_d) \otimes {\cal S}$ will now be lifted to a more complicated complex of vector bundles, and the computation of the hypercohomology involved will be more complicated. More generally one would also like to consider non-polynomial deformations in the fields, as is the case in the derivative expansion of the low energy effective action near the Coulomb branch moduli space of the quantum MSYM theory, where little is known about the constraint from 16 supersymmetries beyond eight-derivative order \cite{Paban:1998qy, Nicolai:2000ht, Maxfield:2012aw}. This is a problem we would like to visit in the future.

An appealing prospective of the on-shell algebraic approach is possibly a classification of higher derivative deformations of maximal supergravity \cite{Cremmer:1978km, Cremmer:1978ds, Brink:1979nt}, in various dimensions (up to 11). In this case, while one can still consider the associative algebra generated by super-gauge covariant derivatives \cite{Brink:1980az,Howe:2003cy,Tsimpis:2004rs}, the relations are not merely quadratic in the generators, and so the machinery of this paper cannot be applied directly. It would be interesting to see whether the cohomology problem of finding nontrivial higher derivative deformations in supergravity can be solved systematically in a purely algebraic approach.

\bigskip

\section*{Acknowledgments}

We are grateful to Shu-Heng Shao for collaboration at the initial stage of the project, and to Clay Cordova, Thomas Dumitrescu, Ken Intriligator, Daniel Jafferis, and Nati Seiberg for helpful discussions. We would like to thank the organizers of the workshop {\it String Geometry and Beyond} at Soltis Center, Costa Rica, the 2013 Simons Workshop in Mathematics and Physics, the KITP program {\it New Methods in Nonperturbative Quantum Field Theory}, and especially the support of KITP during the course of this work. This work is supported in part by a KITP Graduate Fellowship, a Sloan Fellowship, a Simons Investigator Award from the Simons Foundation,  NSF Award PHY-0847457, and by the Fundamental Laws Initiative Fund at Harvard University.

\appendix

\section{Cohomology and deformations}
\label{Lie algebra cohomology}

In this appendix, we recap the notions of Lie algebra cohomology and Hochschild cohomology, and their relation to deformations of an associative algebra, which are standard but perhaps unfamiliar to physicists\footnote{See \cite{deAzcarraga:1998uy,cartan1999homological} for more detailed discussions.}.  Everything introduced here for Lie algebras can be easily generalized for Lie super-algebras.

\subsection{Lie algebra (co)homology}

Let $\cG$ be a Lie algebra and $N$ a $\cG$-module. An $N$-valued $p$-cochain is a skew-symmetric $p$-linear map $c:\cG\wedge\stackrel{p}{\cdots} \wedge\cG\to N.$  The (abelian) group of all $p$-cochains is denoted by ${\rm C}^p(\cG, N)$, i.e., ${\rm C}^p(\cG, N)= \text{Hom}(\Lambda^p\cG,N)$.  The {\it Lie algebra cohomology} ${\rm H}^* (\cG, N)$ is defined as the cohomology of the complex ${\rm C}^* (\cG, N)$ equipped with the coboundary map $d: {\rm C}^{p-1} (\cG, N) \to {\rm C}^p (\cG, N)$ that is defined as follows.  For $c \in {\rm C}^p (\cG, N)$,
\ie\label{Cartan}
% (30) in 06
dc(x_1,\cdots,x_p)= & \sum^p_{i=1} (-1)^{i} x_i\cdot c(x_1,\cdots,\widehat x_i,\cdots,x_p). \\
& + \sum_{1\le i<j\le p}(-1)^{i+j-1}c([x_i,x_j],x_1,\cdots,\widehat x_i,\cdots,\widehat x_j,\cdots,x_p).
\fe

Similarly, the {\it Lie algebra homology} ${\rm H}_* (\cG, N)$ is the homology of the chain complex ${\rm C}_*(\cG, N) \equiv \Lambda^* \cG \otimes N$ with respect to the boundary map $d: {\rm C}_p(\cG, N) \to {\rm C}_{p-1}(\cG, N)$ defined as
\ie\label{lie-hom}
d(x_1 \wedge \dotsb \wedge x_p \otimes m) &= \sum_{i = 1}^p (-1)^i x_1 \wedge \dotsb \wedge \widehat{x_i} \wedge \dotsb \wedge x_p \otimes x_k \cdot m \\
& + \sum_{1 \leq i < j \leq p} (-1)^{i + j } [x_i, x_j] \wedge x_1 \wedge \dotsb \widehat{x_i} \wedge \dotsb \widehat{x_j} \dotsb x_p \otimes m.
\fe

\subsection{Derivations and infinitesimal deformations}

The following Lie algebra cohomology groups have simple interpretations.

\begin{itemize}

\item  {\bf ${\rm H}^1(\cG,\cG)$ as outer derivations.}

A {\it derivation} of a Lie algebra $\cG$ is a linear map $f:\cG\to\cG$ that is compatible with the Lie bracket, i.e.,
\ie
f([a,b])-[f(a),b]-[a,f(b)]=0, \quad \forall a,b\in \cG.
\fe
This condition is the same as the Lie algebra cocycle condition $df(a,b)=0$ if we tregard $\cG$ as a $\cG$-module by action of the Lie bracket.  An {\it inner derivation} of $\cG$ is a linear map $g_x:\cG\to\cG$ such that $g_x:a \mapsto [x,a]$ for a fixed $x \in \cG$.  This may be re-expressed as the coboundary condition $g_x = d x$.  The cohomology classes in ${\rm H}^1(\cG,\cG)$ are called the {\it outer derivations} of $\cG$.

\item {\bf ${\rm H}^2(\cG,\cG)$ as infinitesimal deformations of $\cG$.}

The Lie bracket is a bilinear map $m: \cG\wedge \cG\to \cG$.  Consider an infinitesimal deformation of the Lie bracket from $m$ to $m+\delta m$. In order to preserve the Jacobi identity, we require
\ie
(m+\delta m)((m+\delta m)(a,b),c)+(\text{cyclic permutations}) = 0.
\fe
At linear level in $\delta m$, this condition becomes
\ie
 \delta m([a,b],c) - [c,
\delta m(a,b)]+(\text{cyclic permutations}) \equiv d(\delta m)(a,b,c) = 0.
\fe
Hence consistent deformations of $\cG$ correspond to 2-cocycles of the Lie algebra cohomology with coefficients in $\cG$.   Trivial deformations $\delta m$ are infinitesimal homomorphisms $\delta f:\cG\to \cG$ such that
\ie
(m+\delta m)(a+\delta f (a),b+\delta f (b)) = [a,b]+\delta f([a, b]).
\fe
At linear level in $\delta m$ and $\delta f$, this condition becomes
\ie
 \delta m(a,b) = \delta f([a,b]) - [\delta f(a) ,b] - [a, \delta f(b)] \equiv d(\delta f)(a,b).
\fe
Hence trivial deformations of $\cG$ correspond to 2-coboundaries.  We conclude that nontrivial infinitesimal deformations of $\cG$ are classified by ${\rm H}^2(\cG,\cG)$.

%More generally, we could consider deforming the Lie bracket by elements in a $\cH$, where $\cH$ is a subalgebra of $\cG$.

%We will see in the next section that these deformations can be interpreted as deformations of the associative algebra $U(\cG)$ itself.

\end{itemize}

\subsection{Hochschild cohomology}

The analog of Lie algebra cohomology but for an associative algebra is the {\it Hochschild cohomology}.  For $A$ an associative algebra and $N$ an $A$-bimodule, we have a (Hochschild) complex $\widehat{\rm C}^* (A, N) = \text{Hom}(\otimes^* A, N)$ that is equipped with the differential
\ie\label{Hochschild differential}
dc(x_1, \dotsb, x_{p+1}) & = x_1 \cdot c(x_2, \dotsb, x_{p+1}) + (-1)^{p+1} c(x_1, \dotsb, x_p) \cdot x_{p+1}
\\
& + \sum_{1 \leq i \leq p} (-1)^i c(x_1, \dotsb, x_i x_{i+1}, \dotsb, x_{p+1}).
\fe
The Hochschild cohomology ${\rm HH}^*(A, N)$ is the cohomology of the above complex.  Outer derivations and infinitesimal deformations of $A$ are classified by ${\rm HH}^1 (A, A)$ and ${\rm HH}^2 (A, A)$ (or ${\rm HH}^2 (A, N)$).

%More generally, we could consider deforming the multiplication map by elements in $U(\cH)$, where $\cH$ is a subalgebra of $\cG$.

When $A$ is the universal enveloping algebra $U(\cG)$ of some Lie algebra $\cG$ (isomorphic to $\bigoplus_j {\rm Sym}^j (\cG)$ by the Poincar\'e-Birkhoff-Witt theorem), there is the Cartan-Eilenberg isomorphism ${\rm HH}^* (A, N) \cong {\rm H}^* (\cG, N)$ \cite{cartan1999homological}.

%Such deformations are classified by ${\rm H}^2(\cG,U(\cH))$.  The deformed algebra is no longer a Lie algebra, but as we will see in the next section, these deformations can be interpreted as deformations of the associative algebra $U(\cG)$ itself.

%Therefore, deformations of $\cG$ by $U(\cG)$ or $U(\cH)$ as discussed in the previous section can also be thought of as deformations of the associative algebra $U(\cG)$ itself.

%\item {\bf ${\rm H}^2(\cG,U(\cG))$ as infinitesimal deformations of $U(\cG)$.}

%We can also consider deforming the Lie bracket by elements in its universal enveloping algebra $U(\cG)$, which is isomorphic to $\bigoplus_j {\rm Sym}^j (\cG)$ by the Poincar\'e-Birkhoff-Witt theorem.  As in the previous discussion, nontrivial deformations of $\cG$ by $U(\cG)$ are classified by ${\rm H}^2(\cG, U(\cG))$.  More generally, we could consider deforming the Lie bracket by elements in $U(\cH)$, where $\cH$ is a subalgebra of $\cG$.  Such deformations are classified by ${\rm H}^2(\cG,U(\cH))$.  The deformed algebra is no longer a Lie algebra, but as we will see in the next section, these deformations can be interpreted as deformations of the associative algebra $U(\cG)$ itself.

%

\subsection{Formal deformations}
\label{formal}

A formal deformation of an associative algebra $A$ is a multiplication map $m^t: A\otimes A\to A$ expressed as a formal power series $m^t(a,b) = \sum_{n=0}^\infty m_n(a,b) t^n$, where $m_0(a,b) \equiv ab$ is the undeformed multiplication map.  Associativity requires
\ie
0 &= m^t(m^t(a,b),c) - m^t(a,m^t(b,c))
\\
&= \sum_{i,j=0}^\infty t^{i+j} \left( m_i(m_j(a,b),c) - m_i(a,m_j(b,c)) \right).
\fe
It follows that
\ie
& m_n(ab,c) - m_n(a,bc) + m_n(a,b) c - a m_n(b,c)
\\
& = - \sum_{i=1}^{n-1} \left( m_i(m_j(a,b),c) - m_i(a,m_j(b,c))\right)
\fe
Let us denote that right hand side by $f_n (a, b, c)$.  Since the left hand side takes the form of a coboundary, i.e., $d m_n (a, b, c)$, if $m_n$ were to exist, we need $f_n$ to be a coboundary as well.  In fact, from the definition of $f_n$, one can show that it is always a cocycle, and so the precise requirement is that $f_n$ represents a trivial class in ${\rm HH}^3(A, A)$.

\section{Algebraic notions}
\label{sec:alg-not}

%The main goal of this section is to establish the Koszul duality between certain quadratic associative algebras.  The duality between $U(L)$ and $\cS$ (defined below) is used in the derivation of the long exact sequence (\ref{les}).

In this section we recall a number of algebraic notions relevant to the on-shell formulation of MSYM theories, and basic properties of some of the Lie algebra cohomology groups (these results are due to \cite{Movshev:2003ib,Movshev:2004aw,Movshev:2005ei,Movshev:2009ba}).

\subsection{Some quadratic algebras}

A {\it quadratic algebra} is a graded associative algebra generated by level-1 elements satisfying quadratic (level-2) constraints.  Below are some examples relevant to this paper\footnote{See \cite{polishchuk2005quadratic} for more details.}:
%Let us begin by defining some algebras:
\begin{itemize}
\item  ${\cal S}$ is the algebra of polynomial functions over the space ${\cal C}$ of pure spinors, i.e., polynomials in $\lambda^\A$ subject to 
\ie
\Gamma^m_{\A\B} \lambda^\A \lambda^\B = 0.
\fe
${\cal S}$ can be written as a direct sum $\bigoplus_{k\geq 0}{\cal S}_k$, where each $\cS_k$ is the space of degree-$k$ homogeneous polynomials in $\lambda^\A$.  We can equivalently say that ${\cal S}_k$ is the space of holomorphic sections of the line bundle ${\cal O}(k)$ over the projective pure spinor space ${\cal Q}$.
\item  $B_0$ is the reduced Berkovits algebra, generated by pure spinors $\lambda^\A$ and fermionic spinors $\theta^\A$.
$B_0$ can be regarded as a complex $B_0=\bigoplus_{k\geq 0}(B_0)_k$ equipped with a nilpotent differential $\lambda^\A \partial_{\theta^\A}: (B_0)_k\to (B_0)_{k+1}$, where $k$ is the degree of $\lambda^\A$. There is an odd pairing between $(B_0)_k$ and $(B_0)_{3-k}$ given by
\ie\label{oddpairing}
T_{(\A\B\C) \A_1 \cdots \A_5}\lambda^\A\lambda^\B\lambda^\C\theta^{\A_1}\cdots\theta^{\A_5} \mapsto 1,
\fe
where $T_{(\A\B\C) \A_1 \cdots \A_5}$ is the unique invariant symbol of $SO(10)$ in the above tensor product.

%$3 $\lambda^\A$'s and 5 $\theta^\A$'s.
%
% \item  $B$ is the unreduced Berkovits algebra, generated by $\lambda^\A$, $\theta^\A$ and the coordinates $x^m$ on $\bR^{10}$.  The nilpotent differential is $Q_B = \lambda^\A \left( {\partial \over \partial {\theta^\A}} + \Gamma^m_{\A\B} \theta^\B {\partial \over \partial {x^m}} \right)$.
\end{itemize}

\subsection{Some Lie algebras}

We introduce some Lie super-algebras, including $L$, $YM$ and $YM_d$.
\begin{itemize}

\item  $L$ is the Lie algebra generated by level-1 elements $D_\A$ satisfying
\ie
\Gamma_{mnpqr}^{\A\B} \{ D_\A, D_\B \} = 0.
\fe
$L$ is graded by the level $\bigoplus_{i \geq 1} L^i$.  Its universal enveloping algebra $U(L)$ is a quadratic algebra.

\item  $YM$ is the Lie algebra generated by $D_m$ and $\chi^\A$ subject to the MSYM equations of motion
\ie
& \Gamma^n_{\A\B} [D_n,\chi^\B]=0,
\\
& [D_m, [D_m, D_n] ]+\Gamma^n_{\A\B}\{\chi^\A,\chi^\B\}=0.
\fe
It is isomorphic to $\bigoplus_{i \geq 2} L^i$ via
\ie
& D_m \mapsto {1\over 16} \Gamma_m^{\A\B} \{D_\A, D_\B\}, \quad \chi^\A\mapsto{1\over 10}\Gamma^{\A\B}_m[D_\B,D_m]. 
\fe
We assign a grading for $YM$ according to the grading of $L$.

\item  $YM_d$ is defined as the Lie subalgebra $(L^2)_d \bigoplus_{i \geq 3} L^i$, where $(L^2)_d \equiv \langle D_{d+1}, \dotsc, D_{10} \rangle$ at level-2 can be regarded as the scalars in the $d$-dimensional MSYM theory.  Clearly, $YM_0 \cong YM$.  In \cite{Movshev:2005ei,Movshev:2009ba}, $YM_{10}$ is called $TYM$.

\item  $S$ is the abelian Lie algebra generated by commuting spinors $w_\A$.

\item  $H$ is defined as $L \oplus S$.  It comes with a nilpotent differential $w_\A \partial_{D_\A}:L\to S$ that replaces $D_\A$ by $w_\A$ and acts trivially on $L_n$ for $n\ge 2$.

\end{itemize}

\subsection{Koszul duality}
\label{koszul}

%Two quadratic algebras are {\it quadratic duals} if there is a pairing between their level-1 generators, such that the two sets of constraints are orthogonal complements in the vector space of quadratic constraints.  For super-algebras, 
%
% (for superalgebras, there is a sign that comes into the definition of orthogonality).  For example, the dual of $\cS$ is the universal enveloping algebra $U(L)$, and the dual of $B_0$, denoted by $B_0^!$, is the universal enveloping algebra $U(H)$.

Consider two quadratic super-algebras $\cA$ and $\cA^!$ with level-1 generators $z^i$ and $x_i$, respectively, satisfying the constraints
\ie
r^m_{ij} z^i z^j = 0, \quad s_n^{ij} x^i x^j = 0.
\fe
Suppose there is an odd pairing (parity reversed) between $z^i$ and $x_i$, and the constraints satisfy
\ie
\sum_{i, j} (-1)^{P(z^i) P(x_j)} r^m_{ij} s_n^{ij} = 0,
\fe
where $P$ is the parity.  Then we say that $\cA$ and $\cA^!$ are {\it Koszul duals}.  For example, the quadratic dual of $\cS$ is the universal enveloping algebra $U(L)$, and the quadratic dual of $B_0$, denoted by $B_0^!$, is the universal enveloping algebra $U(H)$.

%Suppose $\cA$ is a graded quadratic algebra with level-1 generators $z^i$, whose Koszul dual $\cA^!$ has level-1 generators $x_i$.

The Koszul complex of $\cA$ is defined as
\ie
K_*(\cA)=\cA\otimes (\cA^!)^*,
\fe
graded by the grading of $\cA$.  $(\cA^!)^*$ is the vector space dual of $\cA^!$.
%which can be view as a right $\cA^!$-module.
%The $\cA$ action on $(\cA^!)^*$ is defined as $(\varphi a)(b)\equiv\varphi(ab)$, where $a\in \cA^!$ acts on $\varphi\in (\cA^!)^*$ and $\varphi a\in (\cA^!)^*$ is evaluated on $b\in (\cA^!)^*$
We define a right $\cA^!$ action $(\cA^!)^* \to (\cA^!)^*, ~ \varphi \mapsto \varphi x$ by $(\varphi x) (y) \equiv \varphi(xy)$, where $x, y \in \cA^!$ and $\varphi \in (\cA^!)^*$.  Then the Koszul complex is an $\cA\otimes \cA^!$-module equipped with the differential $z^i \otimes x_i$, which is nilpotent by orthogonality of the quadratic constraints on $\cA$ and $\cA^!$.  If the Koszul complex is acyclic, then the algebra $\cA$ called a {\it Koszul algebra}.
%, and its {\it Koszul dual} is $\cA^!$.

Now suppose $\cA$ is a Koszul algebra whose dual $\cA^!$ is the universal enveloping algebra of some Lie algebra $\cG$.  From Koszul duality theory \cite{polishchuk2005quadratic}, there are isomorphisms
\ie
&{\rm H}^i (\cG, N) \cong {\rm H}^i (\cA \otimes N, z^i \otimes x_i),
\\
&{\rm H}_i (\cG,N)\cong {\rm H}^{-i} (\cA^* \otimes N, z^i \otimes x_i),
\fe
where $N$ is any $\cG$-module.  For example, we have\footnote{
More explicitly, the isomorphism is induced by a map on the space of cochains ${\rm C}^p (L, N) \to \cS_p \otimes N, ~ c \mapsto \lambda^{\A_1}\cdots\lambda^{\A_p} ~ c(D_{\A_1}\wedge\cdots\wedge D_{\A_p})$.
}
\ie
& {\rm H}^i (L, N) \cong {\rm H}^i (\cS \otimes N, \lambda^\A D_\A), 
\fe
and\footnote{
The $\lambda^\A$ action on $\cS^* \otimes N$ is implemented by first writing the chains in $(\cS^* )_n\otimes N$ in the form $\bar\lambda_{\A_1}\cdots\bar\lambda_{\A_n} \otimes G^{\A_1\cdots \A_n}$, such that $G^{\A_1\cdots \A_n}$ is projected onto the representation $[0,0,0,0,n]$, and then taking the $\bar\lambda_\A$ derivative.
}
\ie
& {\rm H}_i (L, N) \cong {\rm H}^{-i} (\cS^* \otimes N, \lambda^\A D_\A).
\fe

\subsection{Poincar\'e isomorphism}
\label{Poincare isomorphism}

We first establish the isomorphism
\ie\label{ISOB0}
{\rm H}^j (YM, N) \cong {\rm H}^j (B_0 \otimes N, \lambda^\A D_\A + \lambda^\A \partial_{\theta^\A}).
\fe
Let $N$ be an $H$-module on which the action of $w_\A$ is trivial.  Consider the double complex $\bigoplus_{i \leq j} {\rm E}_0^{i,j} = \bigoplus_{i \leq j} \Lambda^{j-i} (L) \otimes \Lambda^i (S) \otimes N$:
\begin{diagram}\label{YMdblcplx}
& & & & \Lambda^0 (L) \otimes \Lambda^2 (S) & \lTo^{d_H} \\
& & & & \uTo^{w_\A \partial_{D_\A}} \\
& & \Lambda^0 (L) \otimes \Lambda^1 (S)  & \lTo^{d_H} & \Lambda^1 (L) \otimes \Lambda^1 (S) & \lTo^{d_H} \\
& & \uTo^{w_\A \partial_{D_\A}} & & \uTo^{w_\A \partial_{D_\A}} \\
\Lambda^0 (L) \otimes \Lambda^0 (S) & \lTo^{d_H} & \Lambda^1 (L) \otimes \Lambda^0 (S)  & \lTo^{d_H} & \Lambda^2 (L) \otimes \Lambda^0 (S) & \lTo^{d_H} \\
\end{diagram}
where $d_H$ is the Lie algebra boundary map ($N$ is omitted in the diagram).\footnote{
Elements of $\Lambda^i (L^1) \otimes \Lambda^j (S)$ can be regarded as polynomials in even variables $D_\A$ and odd variables $w_\A$.  Then $d = w_\A \partial_{D_\A}$ acts as an exterior derivative on the linear space spanned by $D_\A$ ($w_\A = d D_\A$).
%In $\Lambda^i (L^1) \otimes \Lambda^j (S)$, $D_\A$ are regarded as even variables, and $w_\A$ as odd variables.  $w_\A \partial_{D_\A}$ acts as an exterior derivative on the linear space spanned by $D_\A$.
}
Consider the spectral sequence with $d_0 = w_\A \partial_{D_\A}$ and $d_1 = d_H$.  On the first page, the complex collapses to the bottom row\footnote{
This can be phrased as the statement that the inclusion from $YM$, regarded as a one-term complex with trivial differntial, to $(H, w_\A \partial_{D_\A})$ is a quasi-isomorphism.
}
\ie
{\rm E}_1^{i, j} = (\Lambda^j (YM) \otimes N) \delta^i_0.
\fe
The spectral sequence then stabilizes on the second page
\ie
{\rm E}_2^{i, j} = {\rm H}_j (YM, N) \delta^i_0.
\fe
Hence we obtain an isomorphism of homology ${\rm H}_j (YM, N) \cong {\rm H}_j (H, N, d_H + w_\A \partial_{D_\A})$.

%By Koszul duality $U(H) = B_0^!$, there is a further isomorphism\footnote{
%%
%The isomorphism ${\rm H}_* (H, N, d_H + w_\A \partial_{D_\A}) \cong {\rm H}_* (B_0^* \otimes N, \lambda^\A D_\A + \lambda^\A \partial_{\theta^\A})$ is induced by a map on the space of chains $\Lambda^i(L)\otimes \Lambda^j(S) \otimes N \to (B_0)_{i} \otimes N, ~ D_{\A_1} \wedge \cdots \wedge D_{\A_i} \otimes w_{\B_1} \wedge \cdots \wedge w_{B_j} \otimes c \to \tilde\lambda_{\A_1} \cdots \tilde\lambda_{\A_i} \theta_{\B_1} \cdots \theta_{\B_j} \otimes c$.
%%
%\ie\label{ISOB0}
%{\rm H}_* (YM, N) \cong {\rm H}_* (H, N, d_H + Q_H) \cong {\rm H}_* (B_0^* \otimes N, \lambda^\A D_\A + \lambda^\A \partial_{\theta^\A}).
%\fe

A similar analysis gives an isomorphism of cohomology, ${\rm H}^j (YM, N) \cong {\rm H}^j (H, N, d_H + Q_H )$.  Here, $d_H$ denotes the Lie algebra coboundary map, and $Q_H: {\rm C}^{i+j}(\Lambda^i(L)\otimes \Lambda^j(S),N)\to  {\rm C}^{i+j}(\Lambda^{i+1}(L)\otimes \Lambda^{j-1}(S),N)$ is a map induced by
%\footnote{
%%
%More explicitly, given $c \in {\rm C}^{i+j}(\Lambda^i(L)\otimes \Lambda^j(S),N)$, $Q_H$ maps $c$ to $c \circ (w_\A \partial_{D_\A})$.
%}
%
$w_\A \partial_{D_\A}$, $c \mapsto c \circ (w_\A \partial_{D_\A})$.  By Koszul duality $U(H) = B_0^!$, there are further isomorphisms\footnote{
The isomorphism ${\rm H}^* (H, N, d_H + Q_H) \cong {\rm H}^* (B_0 \otimes N, \lambda^\A D_\A + \lambda^\A \partial_{\theta^\A})$ is induced by a map on the space of cochains ${\rm C}^{i+j}(\Lambda^i(L)\otimes \Lambda^j(S),N) \to (B_0)_{i} \otimes N, ~ c \mapsto \lambda^{\A_1}\cdots\lambda^{\A_i}\theta^{\B_1}\cdots\theta^{\B_j} ~ c(D_{\A_1}\cdots D_{\A_i}w_{\B_1}\cdots w_{B_j})$.  The differential $Q_H$ acting on ${\rm C}^*(H, N)$ becomes $\lambda^\A\partial_{\theta^\A}$ acting on $B_0\otimes N$ following from
\ie
&\lambda^{\A_1}\cdots\lambda^{\A_{i+1}}\theta^{\B_1}\cdots\theta^{\B_{j-1}} ~ c(w_\A \partial_{D_\A}(D_{\A_1}\cdots D_{\A_{i+1}}w_{\B_1}\cdots w_{\B_{j-1}}))
\\
&=\lambda^\A\partial_{\theta^\A}\left(\lambda^{\A_1}\cdots\lambda^{\A_i}\theta^{\B_1}\cdots\theta^{\B_j} c(D_{\A_1}\cdots D_{\A_i}w_{\B_1}\cdots w_{\B_j})\right).
\fe
The isomorphism ${\rm H}_* (H, N, d_H + Q_H) \cong {\rm H}_* (B_0^* \otimes N, \lambda^\A D_\A + \lambda^\A \partial_{\theta^\A})$ follows from similar reasoning.
}
\ie\label{ISOB0}
& {\rm H}^i (YM, N) \cong {\rm H}^i (H, N, d_H + Q_H) \cong {\rm H}^i (B_0 \otimes N, \lambda^\A D_\A + \lambda^\A \partial_{\theta^\A}), \\
& {\rm H}_i (YM, N) \cong {\rm H}_i (H, N, d_H + w_\A \partial_{D_\A}) \cong {\rm H}^{-i} (B_0^* \otimes N, \lambda^\A D_\A + \lambda^\A \partial_{\theta^\A}).
\fe

The odd pairing \eqref{oddpairing} on the Berkovits algebra $B_0$ gives an isomorphism between ${\rm H}^i (B_0 \otimes N, \lambda^\A D_\A + \lambda^\A \partial_{\theta^\A})$ and ${\rm H}^{3-i} (B_0^* \otimes N, \lambda^\A D_\A + \lambda^\A \partial_{\theta^\A})$.  The isomorphisms \eqref{ISOB0}, thereby, give Poincar\'e isomorphism
\ie
P: {\rm H}^{i}(YM, N) \stackrel{\sim}{\longrightarrow} {\rm H}_{3-i}(YM,N).
\fe

\subsection{${\rm H}^0$ and ${\rm H}^1$}

The cohomology groups ${\rm H}^n (\cG, {\rm Sym}^j (YM_d))_\ell$ for $\cG = L, YM$ and $n = 0, 1$ can be explicitly computed.  Given the isomophisms
\ie
& {\rm H}^n (L, {\rm Sym}^j (YM_d))_\ell \cong {\rm H}^n(\cS\otimes {\rm Sym}^j (YM_d),\lambda^\A D_\A)_\ell, \\
& {\rm H}^n (YM, {\rm Sym}^j (YM_d))_\ell \cong {\rm H}^n(B_0 \otimes {\rm Sym}^j (YM_d),\lambda^\A D_\A+\lambda^\A\partial_{\theta^\A})_\ell,
\fe
we will present the representatives of the cohomology classes both as elements of the Lie algebra cochains, and as elements in $\cS \otimes {\rm Sym}^j (YM_d)$ or $B_0 \otimes {\rm Sym}^j (YM_d)$.  The results are summarized in Tables \ref{HL}-\ref{HYMd}.  We write $m, n$ for $SO(10)$ indices, $\mu$ for $SO(d)$, and $a, b$ for $SO(10-d)$.

\begin{table}[H]
\centering
\begin{tabular}{|c|c|c|}
\hline
${(}n,j,\ell{)}$ & &   \\
\hline\hline
${(}0,0,0{)}$& $0\mapsto x$ for $x\in\bC$ & ${1}$  \\
\hline
 ${(}1,0,-1{)}$ & $D_\A\mapsto 1$, otherwise $y\mapsto 0$& ${\lambda^\A}$\\
\hline
${(}1,1,1{)}$ & $y\mapsto [D_\A,y]$ & ${(\Gamma^m\lambda)_\A D_m}$\\
\hline
\end{tabular}
\caption{Representatives of classes in ${\rm H}^n(L,{\rm Sym}^j(YM))_\ell$ for $n=0,1$.}
\label{HL}
\end{table}

\begin{table}[H]
\centering
\begin{tabular}{|c|c|c|}
\hline
${(}n,j,\ell{)}$ & &   \\
\hline\hline
${(}0,0,0{)}$& $0\mapsto x$ for $x\in\bC$ & ${1}$  \\
\hline
 ${(}1,0,-1{)}$ & $D_\A\mapsto 1$, otherwise $y\mapsto 0$& ${\lambda^\A}$\\
\hline
${(}1,1,2{)}$ & $y\mapsto [D_\m,y]$ & ${\lambda\Gamma_\m\chi}$\\
\hline
\end{tabular}
\caption{Representatives of classes in ${\rm H}^n(L,{\rm Sym}^j(YM_d))_\ell$ for $n=0, 1$ and $d \geq 1$.}
\label{HLd}
\end{table}

\begin{table}[H]
\centering
\begin{tabular}{|c|c|c|}
\hline
${(}n,j,\ell{)}$ & &   \\
\hline\hline
${(}0,0,0{)}$& $0\mapsto x$ for $x\in\bC$ & ${1}$  \\
\hline
 ${(}1,0,-2{)}$ & $D_m\mapsto 1$, otherwise $y\mapsto 0$& ${\lambda\Gamma^m \theta}$\\
\hline
${(}1,0,-3{)}$  & $\chi^\A\mapsto 1$, otherwise $y\mapsto 0$ & ${ (\theta\Gamma^{mnp}\theta) (\Gamma_{mnp}\lambda)_\A}$ \\
\hline
${(}1,1,0{)}$& $y\mapsto \text{deg}(y)y$ & ${(\lambda\Gamma^m\theta) D_m}$\\
\hline
${(}1,1,0{)}$& $y\mapsto \Lambda_{mn}y$ & ${ (\theta\Gamma_{mnp}\theta) (\lambda\Gamma^{p} \chi)+4(\lambda\Gamma_{{(}m}\theta)D_{n{)}}}$\\
\hline
${(}1,1,1{)}$ & $y\mapsto [D_\A,y]$ & ${ (\Gamma^m\lambda)_\A D_m}$ \\
\hline
\end{tabular}
\caption{Representatives of classes in ${\rm H}^n(YM,{\rm Sym}^j(YM))_\ell$ for $n=0,1$. $\Lambda_{mn}$ are the generators of $SO(10)$ rotations. }
\label{HYM}
\end{table}

\begin{table}[H]
\centering
\begin{tabular}{|c|c|c|}
\hline
${(}n,j,\ell{)}$ & &   \\
\hline\hline
${(}0,0,0{)}$& $0\mapsto x$ for $x\in\bC$ & ${1}$  \\
\hline
 ${(}1,0,-2{)}$ & $D_m\mapsto 1$, otherwise $y\mapsto 0$& ${\lambda\Gamma^m \theta}$\\
\hline
${(}1,0,-3{)}$  & $\chi^\A\mapsto 1$, otherwise $y\mapsto 0$ & ${ (\theta\Gamma^{mnp}\theta) (\Gamma_{mnp}\lambda)_\A}$ 
\\
\hline
${(}1,1,0{)}$& $y\mapsto \Lambda_{ab}y$ & ${ (\theta\Gamma_{abp}\theta) (\lambda\Gamma_{p} \chi)+4(\lambda\Gamma_{{(}a}\theta)D_{b{)}}}$
\\
\hline
${(}1,1,1{)}$ & $y\mapsto [D_\A,y]$ & ${(\lambda\Gamma_m\chi)(\Gamma^m\theta)_\A}$\\
\hline
${(}1,1,2{)}$ & $y\mapsto [D_\m,y]$ & ${\lambda\Gamma_\m\chi}$ \\
\hline
\end{tabular}
\caption{Representatives of classes in ${\rm H}^n(YM,{\rm Sym}^j(YM_d))_\ell$ for $n=0, 1$ and $d \geq 1$. $\Lambda_{ab}$ are the generators of $SO(10-d)$ rotations.}
\label{HYMd}
\end{table}

\section{Kernel of $i^*:{\rm H}^2(L,U(YM_d))\to {\rm H}^2(YM,U(YM_d))$}
\label{Kernel of i^*}

The kernel of $i^*:{\rm H}^2(L,U(YM_d))\to {\rm H}^2(YM,U(YM_d))$ can be studied via the Hochschild-Serre spectral sequence
\ie\label{HSSSE}
&{\rm E}_2^{i,j}\equiv {\rm H}^i(L/YM,{\rm H}^j(YM,U(YM_d)))\Rightarrow {\rm H}^{i+j}(L,U(YM_d)).
\fe
At the infinity page, ${\rm H}^2(L,U(YM_d))$ is isomorphic to the direct sum ${\rm E}^{0,2}_\infty\oplus{\rm E}^{1,1}_\infty\oplus{\rm E}^{2,0}_\infty$. We argue that the space ${\rm E}^{0,2}_\infty$ is isomorphic to the image of the map $i^*$, and ${\rm E}^{1,1}_\infty\oplus{\rm E}^{2,0}_\infty$ is the isomorphic to the kernel of $i^*$.

The inclusion $i:YM\hookrightarrow L$ induces a map from the spectral sequence \eqref{HSSSE} to the spectral sequence
\ie
&\widetilde {\rm E}_2^{i,j}\equiv {\rm H}^i(YM/YM,{\rm H}^j(YM,U(YM_d)))\Rightarrow {\rm H}^{i+j}(YM,U(YM_d)).
\fe
%We will denote all of them by $i^*$, i.e. $i^*:{\rm E}^{j,i}_r \to\widetilde {\rm E}^{j,i}_r$ for every $i,j,r$, and $i^*$ commute with all the differentials $d_r$. 
Since $\widetilde{\rm E}^{1,1}_\infty\oplus\widetilde{\rm E}^{2,0}_\infty$ is trivial due to the triviality of $\widetilde {\rm E}^{1,1}_2\oplus\widetilde {\rm E}^{2,0}_2$, ${\rm E}^{1,1}_\infty\oplus{\rm E}^{2,0}_\infty$ should be inside the kernel of $i^*$.  Furthermore, since the map 
\ie
{\rm E}^{0,2}_2={\rm H}^0(L/YM,{\rm H}^2(YM,U(YM_d)))\to {\rm H}^0(YM/YM,{\rm H}^2(YM,U(YM_d)))
\fe
is an injection, the space ${\rm E}^{0,2}_\infty\subset {\rm E}^{0,2}_2$ should also map injectively into ${\rm H}^{2}(YM,U(YM_d))$. Therefore, ${\rm E}^{0,2}_\infty$ is isomorphic to the image of $i^*$. The kernel of $i^*$ is then isomorphic to ${\rm E}^{1,1}_\infty\oplus{\rm E}^{2,0}_\infty$.

In 0 dimension, from our knowledge of the cohomology groups in Table~\ref{HYM}, we know that the classes inside the kernel of $i^*$ must have dimension $-2,-1$ or $0$, and symmetric power 0 or 1. For even grading, the only two possibilities are $\lambda\Gamma^{mnpqr}\lambda$ and $(\lambda\Gamma^{mnpqr}\lambda)D_r$, expressed in terms of cochains in the complex $\cS\otimes U(YM)$. They are trivial inside ${\rm H}^2(YM,U(YM))$ by $\lambda\Gamma^{mnpqr}\lambda=Q(\lambda\Gamma^{mnpqr}\theta)$ and $\lambda\Gamma^{mnpqi}\lambda D_i=Q(\lambda\Gamma^{mnpqr}\theta D_r+4\lambda\Gamma^{[mnp}\theta D^{q]})$, where $Q = \lambda^\A (D_\A+\partial_{\theta^\A})$ is the differential of the complex $B_0\otimes U(YM)$. We conclude that $\lambda\Gamma^{mnpqr}\lambda$ and $(\lambda\Gamma^{mnpqr}\lambda)D_r$ are the only even classes inside the kernel of $i^*$.

Our analysis can be generalized to dimensions $d \geq 1$. The kernel of $i^*$ in higher dimensions are $\lambda\Gamma^{mnpqr}\lambda$ for all $d$, $(\lambda\Gamma^{1abcd} \lambda)D_a$ for $d=1$, $(\lambda\Gamma^{12abc}\lambda)D_a$ for $d=2$, $(\lambda\Gamma^{123ab}\lambda)D_a$ for $d=3$, and $(\lambda\Gamma^{1234a}\lambda)D_a$ for $d=4$.

\section{Bundles over the projective pure spinor space $\cQ$ and a quasi-isomorphism}
\label{sec:bundles}

Let $\cC$ be the space of pure spinors.  The {\it projective pure spinor space (isotropic Grassmannian)} ${\cal Q}$ is the compact space obtained from the projectivization of ${\cal C} - \{0\}$.  It can be represented as $Spin(10,\mathbb{C}) / P$, where $P \supset GL(5, \bC)$ is the stabilizer of an arbitrarily chosen point on ${\cal Q}$.  Under $SO(10, \bC) \to GL(5, \bC)$, the vector ${\bf 10}$ decomposes into ${\bf 5} \oplus {\bf\overline 5}$, which we denote by $W \oplus W^*$.   $W$ and $W^*$ have charges ${2\over 5}$ and $-{2\over 5}$, respectively, with respect to the diagonal $U(1) \subset GL(5, \bC) \subset P$.  While $W^*$ is a representation of $P$, $W$ is not.\footnote{
%The Lie algebra of $P$ can be understood as follows.
The generators of the Lie algebra $so(10, \mathbb{C})$ decompose into ${\bf adj} + {\bf 10} + {\bf \overline{10}}$, or $W \otimes W^* \oplus \Lambda^2 W \oplus \Lambda^2 W^*$.  The Lie algebra of $P$ is $W \otimes W^* \oplus \Lambda^2 W^*$.  The part $W \otimes W^*$ maps $W \to W$ and $W^* \to W^*$, while $\Lambda^2 W^*$ acts trivially on $W^*$ and maps $W \to W^*$.  Due to this last action, $W$ is not a representation of $P$.
}
For each integer $n$, there is a one-dimensional representation $\mu_n$ of $P$ with charge $n$.  We have $\det W^* = \Lambda^5 W^* \cong \mu_{-2}$ and $\det W = \Lambda^5 W \cong \mu_{2}$.

The representations $W^*$ and $\mu_n$ naturally induce vector bundles over $\cQ$ with structure group $P$ via
\ie
\cW^* = {W^* \times Spin(10, \bC) \over P}, \quad \cO(n) = {\mu_n \times Spin(10, \bC) \over P},
\fe
where $P$ simultaneously acts on the representation $W^*$ (resp. $\mu_n$) and $Spin(10, \bC)$ by,
\ie
p\cdot (g, r)=(gp,\rho(p)^{-1}r),~~{\rm for}~p\in P, ~g\in Spin(10,\mathbb{C}),~r\in R,
\fe
and $(\rho,R)$ corresponds to the representation $W^*$ (resp. $\m_n$).
 
Let $V$ be the vector representation ${\bf 10}$ of $SO(10, \bC)$, and denote the trivial bundle $V \otimes \cQ$ by $\cal V$.  Then $\cW$ is defined as the quotient bundle ${\cal V} / \cW^*$.
% \otimes \cO(-1)$ (the reason for including this $\cO(-1)$ will be clear later).

The bundles $\cW^*$ and $\cW$ also have a more geometric description.  Take the trivial bundle ${\cal V}$.  Given a point $\lambda \in {\cal Q}$, labelled by a pure spinor $\lambda$ up to rescaling, we define $W^*(\lambda)$ to be the subspace of $V$ that annihilates $\lambda$, i.e., spanned by vectors $v_m$ that obey $v_m \Gamma^m_{\A\B} \lambda^\B=0$.  There are 11 independent spinors $\mu$ tangent to the pure spinor space at $\lambda$, such that $\mu\Gamma^m \lambda=0$, so there are only 5 independent constraints on $v_m$.  The subspace $W^*(\lambda)$ is thus 5-dimensional, and defines a rank-5 vector bundle over ${\cal Q}$, which is what we call ${\cal W}^*$.  Similarly, ${\cal W}$ can be defined as the fibration of $(V/W^*(\lambda))$.
% \otimes \mu_{-1}$.

Let us introduce another type of bundles over $\cQ$.  Given a graded $L$-module $N = \bigoplus_n N_n$, let us define a complex
\ie
(N_P)_\ell \equiv \bigoplus_n N_{n+\ell} \otimes \mu_n,
\fe
equipped with the differential $Q = \lambda^\A D_\A$.  This complex can be lifted to a complex of vector bundles over $\cQ$
\ie
\cN_\ell \equiv \bigoplus_n N_{n+\ell} \otimes \cO(n),
\fe
where $N_{n+\ell}$ are trivial bundles, and the differential $Q$ lifts to a differential on $\cN_\ell$ by regarding $\lambda^\A$ as a section of $\cO(1)$ that acts on sections of $\cN_\ell$.

%When $\ell = 0$, we simply write $N_P$ and $\cN$ in place of $(N_P)_0$ and $\cN_0$, respecitvely.

In this paper, we will be considering $N_\ell = {\rm Sym}^j (YM_d)_\ell$ and $\cN_\ell = {\rm Sym}^j ({\cal YM}_d)_\ell$. \\

%We write $({YM_d})_P$ and ${\cal YM}_d$ in place of $({YM_d})_P$ and $({\cal YM}_d)_0$, respecitvely. \\

%The inclusion $YM_d \subset L$ naturally defines for us the chain complexes $(YM_d)_P$ and $({\cal YM}_d)_\ell$.

\noindent{\bf A quasi-isomorphism}

The complex $YM_P \otimes \mu_{-2}$ decomposes with respect to representations of $P$ as
\ie
%& L^1 \otimes \mu_{-1} \to \mu_{-2} \oplus \Lambda^3 W^* \oplus W^*, \\
& L^2 \otimes \mu_0 \to W \oplus W^*, \\
& L^3 \otimes \mu_1 \to W \oplus \Lambda^2 W^* \otimes \mu_2 \oplus \mu_2, \\
& L^4 \otimes \mu_2 \to W \otimes W^* \otimes \mu_2 \oplus \Lambda^2 W \otimes \m_2 \oplus \Lambda^2 W^* \otimes \mu_2, \\
& \dotsb
\fe
In \cite{Movshev:2009ba} it is shown that $W^* \hookrightarrow L^2 \otimes \mu_0 \subset (YM_P \otimes \mu_{-2}, Q)$ is in fact a quasi-isomorphism.  Namely, $Q$ acts by
\footnote{ Consider the first map.  Given an element $v^m D_m \in L^2 \cong V$, the $Q$ action on it gives $v^m (\lambda \Gamma_m \chi)$.  The condition $v^m \Gamma^m_{\A\B} \lambda^\B = 0$ precisely defines the subspace $W^* \subset V$, and therefore $W^* \mapsto 0$. }
\ie
L^2 \otimes \mu_0 \to L^3 \otimes \mu_1: \quad & W \mapsto W, \quad W^* \mapsto 0 \\
L^3 \otimes \mu_1 \to L^4 \otimes \mu_2: \quad & W \mapsto 0, \quad \Lambda^2 W^* \otimes \mu_2 \mapsto \Lambda^2 W^* \otimes \mu_2, \\
& \mu_2 \mapsto \mu_2 \subset W \otimes W^* \otimes \mu_2, \\
\dotsb
\fe
For $(YM_d)_P$, the difference from $YM_P$ is that $(L^2)_d$, unlike $L^2$, is not a representation of $P$.  However, $(L^2)_d \to W$ is still the only map that can give rise to nontrivial cohomology.  Thus $( (L^2)_d \to W ) \hookrightarrow (YM_d)_P \otimes \mu_{-2}$, where both are equipped with the differential $Q$, is a quasi-isomorphism.  It lifts to a quasi-isomorphism of bundles
\ie\label{quasi-bundle}
( (L^2)_d \to {\cal W} ) \otimes \cO(2) \hookrightarrow {\cal YM}_d.
\fe
By K\"unneth theorem, we can generalize this to quasi-isomorphisms of tensor products of bundles.

\section{A long exact sequence}
\label{sec:les}

The purpose of this section is to review the following long exact sequence derived in \cite{Movshev:2005ei,Movshev:2009ba}
\ie
\cdots & \to {\bf H}^i (\cQ,{\rm Sym}^j ({\cal YM}_d) )_\ell \stackrel{\iota_{2-i}}{\to} {\rm H}_{2-i} (L,{\rm Sym}^j (YM_d) )_{\ell-8} \stackrel{\delta_{i+1} }{\to} {\rm H}^{i+1} (L,{\rm Sym}^j (YM_d) )_\ell
\\
& \to {\bf H}^{i+1} (\cQ,{\rm Sym}^j ({\cal YM}_d) )_\ell \stackrel{\iota_{3-i} }{\to} \cdots
\label{les}
\fe
In the following, we will set
\ie
N_\ell \equiv {\rm Sym}^j (YM_d)_\ell, \quad \cN_\ell \equiv {\rm Sym}^j (\mathcal{YM}_d)_\ell,
\fe
and abbreviate $\otimes \cO(n)$ as $(n)$.
%Boldface ${\bf H}$ denotes the hypercohomology which we introduce below.  
%We introduce the shorthand notation
%\ie
%& N_\ell \equiv {\rm Sym}^j (YM_d)_\ell \\
%& \cN_\ell \equiv {\rm Sym}^j (\mathcal{YM}_d)_\ell.
%\fe
%The bundle $\cN_\ell$ can be further filtered into $\bigoplus_n \cN_{\ell, n} \equiv \bigoplus_n N_{\ell + n} \otimes \cO(n)$, where $N_{\ell + n}$ are trivial bundles.

Let us consider the double complex $\bigoplus_{n, a} E_0^{n, a} = \bigoplus_{n, a} \Omega^a ( N_{\ell+n} (n) )$ of $a$-forms valued in $N_{\ell+n} (n)$
\begin{diagram}\label{dblcplx}
& \uTo & & \uTo  & & \uTo
\\
\rTo^{\hspace{.5in}} & \Omega^{2} ( N_{\ell+n} (n) ) & \rTo^Q & \Omega^{2}( N_{\ell+n+1} (n+1) ) & \rTo^Q & \Omega^{2}( N_{\ell+n+2} (n+2) ) & \rTo
\\
& \uTo^{\bar\partial} & & \uTo^{\bar\partial} & & \uTo^{\bar\partial} 
\\
\rTo^{\hspace{.5in}} & \Omega^{1}( N_{\ell+n} (n) ) & \rTo^Q & \Omega^{1}( N_{\ell+n+1} (n+1) ) & \rTo^Q & \Omega^{1}( N_{\ell+n+2} (n+2) ) & \rTo
\\
& \uTo^{\bar\partial} & & \uTo^{\bar\partial} & & \uTo^{\bar\partial} 
\\
\rTo^{\hspace{.5in}} & \Omega^0 ( N_{\ell+n} (n) ) & \rTo^Q & \Omega^0 ( N_{\ell+n+1} (n+1) ) & \rTo^Q & \Omega^0 ( N_{\ell+n+2} (n+2) ) & \rTo
\end{diagram}
%
%The vertical map is $d \equiv \lambda^\A D_\A: \cN_{\ell, n} \to \cN_{\ell, n+1}$, which we can interpret as the differential of a chain complex.  
The vertical map is the Dolbeault operator $\bar\partial: \Omega^a ( N_{\ell+n} (n) ) \to \Omega^{a+1} ( N_{\ell+n} (n) )$, and the horizontal map is $Q \equiv \lambda^\A D_\A: \Omega^a ( N_{\ell+n} (n) ) \to \Omega^a ( N_{\ell+n+1} (n+1) )$, where $\lambda^\A$ is regarded as a section of $\cO(1)$.  The hypercohomology is defined as the cohomology with respect to $\bar\partial + Q$ and is denoted by ${\bf H}^* (\cQ, \cN)_\ell$; the $m$-th hypercohomology group is the direct sum of all $(\bar\partial + Q)$-cohomology classes with ${n} + a = m$.

Now let us consider the spectral sequence for this double complex with $d_0 = \bar\partial$ and $d_1 = Q$.  %The zeroth page is given by $E_0^{k, a} = \Omega^a (\cN_{\ell, k})$.
%The crucial statement is that the hypercohomology is given by the spectral sequence at infinity page:
%\ie
%{\bf H}^m (\cQ, \cN_\ell) = {{}}\bigoplus_{k+a = m} E_\infty^{k, a}.
%\label{hyp2inf}
%\fe
%We will now see that this spectral sequence simplifies dramatically already on the first page.
Because $N_\ell$ is a trivial bundle, on the first page we have $E_1^{{n}, a} = {\rm H}^a (\cQ, N_{\ell+n} (n) ) = N_{\ell+{n}} \otimes {\rm H}^a (\cQ, \cO({n}))$.  The Dolbeault cohomology of $\cO({n})$ can be computed using Borel-Weil-Bott theory.  The only non-vanishing groups are
\ie
& {\rm H}^0 ({\cal Q}, \cO({n})) = [0, 0, 0, 0, {n}] \equiv \cS_{n}, \quad {n} \geq 0,
\\
& {\rm H}^{10} ({\cal Q}, \cO({n})) = [0, 0, 0, -8-{n}, 0] \equiv \cS^*_{-8-{n}}, \quad {n} \leq -8.
\label{bwblist3}
\fe
%For reasons that will become clear later, it is convenient to define
The first page becomes
\begin{diagram}[width=1.5em]
N_{0} \otimes \cS^*_{\ell-8} & \rTo^{d_1} & \cdots & \rTo^{d_1} & N_{\ell-8} \otimes \cS^*_0 \\
&&&&&&&& N_{\ell} \otimes \cS_0 & \rTo^{d_1} & N_{\ell+1} \otimes \cS_1 & \rTo^{d_1} & \cdots
\end{diagram}
where $N_{\ell} \otimes \cS_0$ and $N_{\ell-8} \otimes \cS^*_0$ are located at $(k, a) = (0, 0)$ and $(-8, 10)$, respectively.  Let us define
\ie
& {{}} (N_c)_\ell \equiv \bigoplus_{{n} \geq 0} N_{\ell + {n}} \otimes \cS_{n}, \\
& {{}} (N_h)_\ell \equiv \bigoplus_{-\ell \leq n \leq 0} N_{\ell + n} \otimes \cS^*_{- {n}}.
%& {{}} (N_h)_\ell \equiv \bigoplus_{0 \leq {n} \leq \ell} N_{n} \otimes \cS^*_{\ell - {n}}.
\fe
Since the $d_2, \dotsc, d_{10}$ maps act trivially, we go directly to the eleventh page
\begin{diagram}[2em]
\cdots\quad & {\rm H}^{-3} (N_h)_{\ell-8} & \quad\cdots\quad & {\rm H}^{0} (N_h)_{\ell-8} \\
&& \rdTo(4,2)^{d_{11}^{(0)}} && \rdTo(4,2)^{d_{11}^{(3)}}
\\
&&&& \hspace{.5in} & {\rm H}^0 (N_c)_\ell & ~~\quad\cdots\quad~~ & {\rm H}^3 (N_c)_\ell & \quad \cdots
\end{diagram}
where the only nontrivial $d_{11}$ maps are $d_{11}^{(i)}$ for $i = 0, \dotsc, 3$.  The spectral sequence stabilizes on the twelfth page, therefore
\ie
{\bf H}^m (\cQ, \cN)_\ell = {\rm coker\,} d_{11}^{(m)} \oplus \ker d_{11}^{(m+1)}, \quad m = -1, \dotsc, 3.
\fe
This can be recast into a long exact sequence
\ie
\cdots \to {\bf H}^i (\cQ, \cN)_\ell \to {\rm H}^{i-2} (N_h)_{\ell-8} \to {\rm H}^{i+1} (N_c)_\ell \to {\bf H}^{i+1} (\cQ, \cN)_\ell \to \cdots
\fe
Finally, Koszul duality between $U(L)$ and $\cS$ gives the isomorphisms
\ie
{\rm H}^i (L, N)_\ell \cong {\rm H}^i (N_c)_\ell, \quad {\rm H}_i (L, N)_\ell \cong {\rm H}^{-i} (N_h)_\ell.
\fe
The derivation is now complete. \\

\noindent {\bf A corollary: }  {\it For $\ell > 2$, $\iota_1$ is an injection and $\iota_i$ is an isomorphism for $i \leq 0$.}

This follows from the long exact sequence (\ref{les}) together with our explicit knowledge of ${\rm H}^0 (L, {\rm Sym}^j (YM_d))_\ell$ and ${\rm H}^1 (L, {\rm Sym}^j (YM_d))_\ell$ (Tables~\ref{HL} and~\ref{HLd}). \\

\noindent {\bf Another corollary: }  {\it For $2j - \ell > -8$, ${\rm H}^i (L, {\rm Sym}^j (YM_d) )_\ell \to {\bf H}^i (\cQ,{\rm Sym}^j ({\cal YM}_d) )_\ell$ is an isomorphism.}

This follows from the fact that ${\rm H}_* (L, {\rm Sym}^j (YM_d) )_{\ell - 8 < 2j} \cong 0$.

\section{ SUSY homology and exceptional D-terms}
\label{susyhom}

In this section, following \cite{Movshev:2005ei,Movshev:2009ba}, we show that exceptional D-terms, coming from classes in the cokernel of $B_L: {\rm H_0} (L, {\rm Sym}^{j+1} (YM_d) ) \stackrel{B_{YM}}{\to} B_L: {\rm H_1} (L, {\rm Sym}^j (YM_d) )$, lie inside the SUSY homology.

Consider the double complex $\bigoplus_{i \geq j} E_0^{i,j} = \bigoplus_{i \leq j} \Lambda^{i-j} (L) \otimes {\rm Sym}^j (YM_d)$:
\begin{diagram}\label{susy-dblcplx}
& & & & \Lambda^0 (L) \otimes {\rm Sym}^2 (YM_d) & \lTo^{d_L} \\
& & & & \dTo^{d_{dR}} \\
& & \Lambda^0 (L) \otimes {\rm Sym}^1 (YM_d) & \lTo^{d_L} & \Lambda^1 (L) \otimes {\rm Sym}^1 (YM_d) & \lTo^{d_L} \\
& & \dTo^{d_{dR}} & & \dTo^{d_{dR}} \\
\Lambda^0 (L) \otimes {\rm Sym}^0 (YM_d) & \lTo^{d_L} & \Lambda^1 (L) \otimes {\rm Sym}^0 (YM_d) & \lTo^{d_L} & \Lambda^2 (L) \otimes {\rm Sym}^0 (YM_d) & \lTo^{d_L} \\
\end{diagram}
The horizontal map $d_L$ is the Lie algebra differential (defined in (\ref{lie-hom})), and the vertical map $d_{dR}$ is the de Rham map induced by the inclusion $YM_d \hookrightarrow L$.

The spectral sequence with $d_0 = d_{dR}$ and $d_1 = d_L$ stabilizes already on the second page (due to the same reasons as for (\ref{YMdblcplx})), which is given by
\ie
E_2^{i, j} = {\rm H}_i ({\bf susy}_d) \delta^j_0.
\fe
On the other hand, the first page of the spectral sequence with the opposite choice $d_0 = d_L$ and $d_1 = d_{dR}$ is given by $E_1^{i, j} = {\rm H}_{i-j} ( L, {\rm Sym}^j (YM_d) )$:
%: \Lambda^{j-i} (L) \otimes {\rm Sym}^i (YM_d) \to \Lambda^{j-i-1} (L) \otimes {\rm Sym}^i (YM_d)$ defined as
%\ie
%d(a_0 da_1 \dotsb da_j) &= \sum_{k = 1}^j (-)^k [a_k, a_0] da_1 \dotsb \widehat{da_k} \dotsb da_j \\
%& \hspace{.5in} + \sum_{k, l = 1}^j (-)^{k + l - 1} a_0 d[a_k, a_l] da_1 \dotsb \widehat{da_k} \dotsb \widehat{da_l} \dotsb da_j.
%\fe
%Here (anti-)symmetrization is implicit and $\widehat \quad$ indicates the omission of an element.
\begin{diagram}
& & & & & & {\rm H}_0(L , \text{Sym}^3(YM_d) )
\\
& & & & & & \dTo^{d_1} 
\\
& & & & {\rm H}_0(L , \text{Sym}^2(YM_d) )& & {\rm H}_1(L , \text{Sym}^2(YM_d) )
\\
 & & & & \dTo^{d_1} &\rdDashto(2,4)^{d_2} & \dTo^{d_1} 
\\
& & {\rm H}_0(L , \text{Sym}^1(YM_d) )& &{\rm H}_1(L , \text{Sym}^1(YM_d)) & & {\rm H}_2(L , \text{Sym}^1(YM_d))
\\
 & & \dTo^{d_1} & & \dTo^{d_1} & & \dTo^{d_1} 
\\
{\rm H}_0(L , \text{Sym}^0(YM_d)) & \quad & {\rm H}_1(L , \text{Sym}^0(YM_d) )& \quad & {\rm H}_2(L , \text{Sym}^0(YM_d))  & \quad\quad & {\rm H}_3(L , \text{Sym}^0(YM_d))
\\
&
\end{diagram}

\noindent{\bf The Connes differential}

The de Rham map $d_{dR}: {\rm H}_0 (L, \text{Sym}^{j+1} (YM_d) ) \to {\rm H}_1(L, \text{Sym}^{j} (YM_d) )$ can be identified with the Connes differential $B_L$.  For our purpose, we can simply take this as the definition of $B_L$.  %One can consider a double complex similar to (\ref{susy-dblcplx}) but with Lie algebra $YM$ replacing $L$.  On the first page, 
One can similarly  define the de Rham map $d_{dR}: {\rm H}_0 (YM, \text{Sym}^{j+1} (YM_d) ) \to {\rm H}_1(YM, \text{Sym}^{j} (YM_d) )$ and identify it with the Connes differential $B_{YM}$.\footnote{
If we vary the Lagrangian with respect to a component field $X$, we get $\delta X {\delta \cL \over \delta X}$, where $\delta X \in YM$ and ${\delta \cL \over \delta X} \in U(YM_d)$.  Therefore, $B_{YM}$ can be regarded as varying the Lagrangian to obtain the equations of motion for component fields.
}
\\

To proceed we collect a few ingredients.  First is the fact that $\iota_{i \leq 0}$ are isomorphisms, as explained at the end of Appendix~\ref{sec:les}.\footnote{
Note that $\ell \geq 8$; otherwise ${\rm H}_{\ell - 8}$ is trivial.
}
Another key is that 
the image of $\iota_{i \leq 0}$ survives to the infinity page of the spectral sequence which we shall elucidate below.  It then follows that the $d_{\geq 2}$ maps are trivial, and the spectral sequence stabilizes at the second page. \\

\noindent{\bf Survival of ${\rm im}\,\iota_{i \leq 0}$}

The inclusion $YM_d \hookrightarrow L$ also induces the de Rham map on the Lie algebra cohomology, which fits into the commutative diagram
\begin{diagram}
{\bf H}^{2-i}({\cal Q},{\rm Sym}^j({\mathcal YM}_d))_\ell& \rTo^{\iota_i} &  {\rm H}_{i}(L,{\rm Sym}^j (YM_d))& \rTo^{\delta} &{\rm H}^{3-i}(L,{\rm Sym}^j (YM_d))
\\
& & \dTo^{d_{dR}} & & \dTo^{d_{dR}} 
\\
{\bf H}^{1-i}({\cal Q},{\rm Sym}^{j-1}({\mathcal YM}_d))_\ell& \rTo^{\iota_{i+1}} &  {\rm H}_{i+1}(L,{\rm Sym}^{j-1} (YM_d))& \rTo^{\delta} &{\rm H}^{2-i}(L,{\rm Sym}^{j-1} (YM_d))
\end{diagram}
The commutativity of the diagram implies that the image of $d_{dR} \circ \iota_i$ is inside the image of $\iota_{i+1}$.

Starting with a cycle $a_0\in \Lambda^i(L)\otimes{\rm Sym}^j(YM_d)$ representing a nontrivial class $[a_0]$ in the image of $\iota_i$, $d_{dR} (a_0)$ is a cycle that represents a class in the image of $\iota_{i+1}$ in $ {\rm H}_{i+1}(L,{\rm Sym}^{j-1} (YM_d))$. From the representation content of the hypercohomology, $[d_{dR}(a_0)]$ should be a trivial class; hence, there exists $a_1\in \Lambda^{i+2}(L)\otimes{\rm Sym}^{j-1}(YM_d)$, such that $d_L(a_1)=d_{dR}(a_0)$. Iterating this procedure, we obtain a finite sequence $(a_0,a_1,\cdots, a_n)$, where $a_k\in \Lambda^{i+2k}(L)\otimes{\rm Sym}^{j-k}(YM_d)$ and $d_{dR}(a_n)=0$. From this sequence, we construct a cycle $D=a_0-a_1+a_2+\cdots +(-1)^j a_n$ of the diagonal homology ${\rm H}_{i+2j}( \Lambda(L)\otimes{\rm Sym}(YM_d),d_L+d_{dR})$, which is isomorphic to ${\rm H}_{i+2j}({\bf susy}_d)$. 

Next, we show that the cycle $D$ is nontrivial. The pairing between ${\rm C}^*(L,\mathbb{C})$ and ${\rm C}_*(L,\mathbb{C})$ gives rise to a natural action of ${\rm C}^k(L,\mathbb{C})$ on ${\rm C}_i(L,U(YM_d))$ that is compatible with $d_L$ and commutes with $d_{dR}$, thus inducing an action of ${\rm H}^k(L,\mathbb{C})$ on ${\rm H}_i(L,U(YM_d))$. This is most obvious in terms of the deformation complex ${\rm Sym} (YM_d) \otimes\mathcal{S}^* $, where the generators of ${\rm H}^i(L,\mathbb{C})$ which are degree $i$ monomials in $\lambda^\A$, acts on the chains by multiplication defined by $\la\lambda^\A,\bar\lambda_\B\ra=\delta^\A_\B$. A crucial property of this multiplicative action is that at each degree $i$, the generators of $ {\rm H}^i(L,\mathbb{C})$ have no common kernel.
%from the representation content of the hypercohomology, it is clear that one can augment $a$ by a finite sequence of nonzero elements $D=(a_0=a,a_1,\dots,a_n)$ from the double complex, representing a cocycle of the diagonal cohomology, i.e. in $H_*({\bf susy}_d)$. 
We can choose a generator $f(\lambda)$ of ${\rm H}^i(L,\mathbb{C})$ that maps $[a_0]$ nontrivially to $[a'_0]$ in ${\rm H}_0(L,U(YM_d))$, and again lift $[a'_0]$ to a cocycle of ${\rm H}_*({\bf susy}_d)$ represented by $D'=a'_0-a'_1+a'_2+\dots+(-1)^ma'_m$, which is nontrivial due to the triangular shape of the double complex. Since the multiplicative action commutes with the differentials of the spectral sequence, $D$ and $D'$ are related by $a'_k=f(\lambda)\cdot a_k$ and $m=n$. Therefore, the nontriviality of $D$ in ${\rm H}_*({\bf susy}_d)$ is demanded by that of $D'$. \\

%By Proposition~59 of the 06paper, the image of $\iota_{i \leq 0}$ survive to the infinity page of the spectral sequence, and therefore we know that the $d_{\geq 2}$ maps are trivial, and the spectral sequence stables on the second page. 

The cokernel of $B_L$ modded out by the image of $\iota_1$ precisely classifies the exceptional D-term deformations.  From the preceeding discussion, and the fact that the diagonal cohomology for the two different choices of $d_0$ and $d_1$ are the same, we conclude that
\ie
\label{susyhyper}
{\rm H}_{1+2j} ({\bf susy}_d, \bC)_\ell \cong ({\rm coker}\,B_L/{\rm im}\,\iota_1) \bigoplus_{i+2j' = 1+2j}  {\bf H}^{2-i} (\cQ, {\rm Sym}^{j'} (\mathcal{YM}_d))_\ell.
\fe
Thus knowledge of ${\rm H}_n ({\bf susy}_d, \bC)_\ell$ for odd $n$ and even $\ell$ (odd $\ell$ violates the boson/fermion $\bZ_2$-grading), and ${\bf H}^n (\cQ, {\rm Sym}^j (\mathcal{YM}_d))_\ell$ for $n$ odd and $\leq 1$ is all that is needed to classify exceptional D-term deformations. \\

\noindent{\bf List of SUSY (co)homology in general dimensions}

The ${\bf susy}_d$ cohomology groups were computed in \cite{Movshev:2011pr}.  We list the results in Table~\ref{tab:susyhom}, organized by whether a representation is Lorentz or R-symmetry invariant, or both.  The ${\bf susy}_d$ homology groups can be obtained via the isomorphism
\ie
{\rm H}^{\ell, n} \equiv {\rm H}^n ({\bf susy}, \bC)_\ell \cong {\rm H}_{\ell-n} ({\bf susy}, \bC)_\ell \equiv {\rm H}_{\ell-n, \ell}.
\fe
This isomorphism exchanges the chiral and antichiral representations.

Our convention for the $SO(d)$ Dynkin labels is as follows.  In $d \geq 5$, the leftmost label is the vector.  In $d = 6, 8, 10$, the rightmost is the chiral spinor (eg., $D_\A$), and the second rightmost is the antichiral spinor (eg., $\chi^\A$).  In $d = 4$, the left is the chiral spinor, and the right is the antichiral.  In $d = 2$, the label is the $U(1)$ charge.

\begin{table}[H]
\centering
\begin{tabular}{|c|c|c|c|}
\hline
$d$ & Lorentz $+$ R & Lorentz $-$ R & R $-$ Lorentz \\\hline\hline
all & ${\rm H}^{0, 0}$ && \\\hline
10 & ${\rm H}^{4, 1}$ && ${\rm H}^{k, 0} = [0, 0, 0, 0, k]$ \\
& ${\rm H}^{12, 5}$ && ${\rm H}^{k, 1} = [0, 0, 0, 1, k-3]$ \\
&&& ${\rm H}^{k, 2} = [0, 0, 1, 0, k-6]$ \\
&&& ${\rm H}^{k, 3} = [0, 1, 0, 0, k-8]$ \\
&&& ${\rm H}^{k, 4} = [1, 0, 0, 0, k-10]$ \\
&&& ${\rm H}^{k, 5} = [0, 0, 0, 0, k-12]$ \\\hline
9 & ${\rm H}^{2, 0}$ & & ${\rm H}^{k, 0} = [0, 0, 0, k]$ \\
& $ {\rm H}^{10, 4}$ && ${\rm H}^{k, 1} = [0, 0, 1, k-4]$ \\
&&& ${\rm H}^{k, 2} = [0, 1, 0, k-6]$ \\
&&& ${\rm H}^{k, 3} = [1, 0, 0, k-8]$ \\
&&& ${\rm H}^{k, 4} = [0, 0, 0, k-10]$ \\\hline
8 & ${\rm H}^{8, 3}$ & ${\rm H}^{2k, 0} = [\pm 2k]$ & ${\rm H}^{2k, 0} = [0, 0, k, k]$ \\
&&& ${\rm H}^{2k, 1} = [0, 1, k-2, k-2]$ \\
&&& ${\rm H}^{2k, 2} = [1, 0, k-3, k-3]$ \\
&&& ${\rm H}^{2k, 3} = [0, 0, k-4, k-4]$ \\\hline
7 & ${\rm H}^{6, 2}$ & ${\rm H}^{2k, 0} = [\pm 2k]$ & ${\rm H}^{2k, 0} = [0, k, 0]$ \\
&&& ${\rm H}^{2k, 1} = [1, k-2, 0]$ \\
&&& ${\rm H}^{2k, 2} = [0, k-3, 0]$ \\\hline
6 & ${\rm H}^{4, 1}$ & ${\rm H}^{2k, 0} = [k, k]$ & ${\rm H}^{2k, 0} = [k, 0, 0]$ \\
&&& ${\rm H}^{2k, 1} = [k-2, 0, 0]$ \\\hline
5 & ${\rm H}^{2k, 0}$ & ${\rm H}^{2k, 0} = 2 [k, 0]$ & \\\hline
4 & & ${\rm H}^{2k, 0} = (k+1) [0, k, k]$ & \\\hline
3 & & ${\rm H}^{2k, 0} = \bigoplus_{i=0}^k [i, k-i, 0]$ & \\\hline
2 & & ${\rm H}^{2k, 0} = \bigoplus_{i=0}^{k} [i, 0, k-i, k-i]$ & \\\hline
1 & & ${\rm H}^{k, 0} = \bigoplus_{i=0}^{[k/2]} [i, 0, 0, k-2i]$ & \\\hline
0 & & ${\rm H}^{k, 0} = \bigoplus_{i=0}^{[k/2]} [i, 0, 0, k-2i, 0]$ & \\\hline
\end{tabular}
\caption{Classes in ${\rm H}^{\ell, n} \equiv {\rm H}^n ({\bf susy}, \bC)_\ell$ (this is the notation used in~\cite{Movshev:2011pr}).  The numbers in brackets are Dynkin labels of the corresponding $SO(10-d)$ or $SO(d)$ irrep.}
\label{tab:susyhom}
\end{table}

\section{Computation of hypercohomology}
\label{Hypercohomology}

To classify F-term deformations, we need to know ${\bf H}^2 ( \cQ, {\rm Sym}^j (\mathcal{YM}_d) )_\ell$; to classify exceptional D-term deformations, we need to know ${\bf H}^m ( \cQ, {\rm Sym}^j (\mathcal{YM}_d) )_\ell$ for $m$ odd and $\leq 1$.  In this section, we describe the machinery for explicit computation of these hypercohomology groups, and present explicit results for classes that preserve Lorentz or R-symmetry, or both.

We make use of the quasi-isomorphism (\ref{quasi-bundle})
\ie
( (L^2)_d \to \cW ) \otimes \cO(2) \hookrightarrow \mathcal{YM}_d,
\fe
which induces a quasi-isomorphism from the double complex
\ie\label{quasi-dblcplx}
\bigoplus_{k, a} E_0^{2j - \ell + k, a} = \bigoplus_{k, a} \Omega^a ( {\rm Sym}^{j-k} (L^2)_d \otimes \Lambda^k \cW (2j - \ell) )
\fe
to the double complex
\footnote{
The bundle $\cW$ is embedded into $L^3 \otimes \cO(1)$, so $m$, the degree of the line bundle in ${\rm Sym}^{j-k} (L^2)_d \otimes \Lambda^k \cW (2j - \ell)$, is $0 + k + (2j - \ell)$.
%$\cW$ is embedded in $L^3 \otimes \cO(1)$
%Note that $n$ is defined as ${\rm Sym}^j (YM_d)_{\ell + n} \otimes \cO(n)$, whereas the bundle ${\rm Sym}^j ({\cal YM}_d)$ is the lift of ${\rm Sym}^j (YM_d)_P$.
%The relation $n = 2j - \ell + k$ comes from the following.  Each $({\cal L}^2)_d \subset $ 
%Recall that $n$ is the total degree of the line bundle in ${\rm Sym}^j ({\cal YM}_d)_\ell$.  Since $({\cal L}^2)_d$ contains $\cO(2)$ and $\cW$ contains $\cO(3)$, 
%Again, each $\cW$ contains an $\cO(1)$, so the total is $\cO(2j - \ell + k)$ corresponding to $k' = 2j - \ell + i$ in $({\rm Sym}^j (\mathcal{YM})_d)_{\ell, k'}$.
}
\ie
\bigoplus_{m, a} E_0^{m, a} = \bigoplus_{m, a} \Omega^a ({\rm Sym}^j (YM_d)_{\ell+m} (m) )
\fe
of Appendix~\ref{sec:les}.  As in Appendix~\ref{sec:les}, $\otimes \cO(2j - \ell)$ is abbreviated as $(2j - \ell)$.  Let us now define $n \equiv 2j - \ell$.  The symbol $n$ will be reserved for this defition throughout the rest of this section.

%The double complex (\ref{quasi-dblcplx}) is

%Here $(2j - \ell)$ is the abbreviation of $\otimes \cO({n})$.  This is the notation we will use throughout this section.
%We define $q \equiv 2j - \ell$, and for the rest of this section we will speak of $q$ instead of $\ell$.

%\ie
%\bigoplus_{k, a} E_0^{2j - \ell + k, a} = \bigoplus_{k, a} \Omega^a ( {\rm Sym}^{j-k} (\cL^2)_d \otimes \Lambda^k \cW (2j - \ell) )
%\fe

Consider the spectral sequence of this latter complex with $d_0 = \bar\partial$ and $d_1 = Q$.  Since $(L^2)_d$ is a trivial bundle, on the first page we just have
\ie
E_1^{n + k, a} = {\rm Sym}^{j-k} (L^2)_d \otimes {\rm H}^a ( \cQ, \Lambda^k \cW (n) ).
\fe
The hypercohomology, which is the same for the two quasi-isomorphic complexes, is related to the infinity page of this spectral sequence by
\ie
{\bf H}^{2j - \ell + m} (\cQ, {\rm Sym}^j (\mathcal{YM}_d) )_\ell \cong {{}} \bigoplus_{k+a = m} E_\infty^{2j - \ell + k, a}.
\label{hyper2inf}
\fe
%After taking the cohomology of the Dolbeault differential $d_0 = \bar\partial$, we get on the first page $E_1^{2j - \ell + a, k} = Sym^{j-k} (L^2)_d \otimes {\rm H}^a ( \cQ, \Lambda^k \cW (2j - \ell) )$ since $(\cL^2)_d$ is a trivial bundle.  The hypercohomology is related to the infinity page of the spectral sequence by
%\ie
%{\bf H}^{2j - \ell + m} (\cQ, {\rm Sym}^j \mathcal{YM}_d)_\ell \cong {{}} \bigoplus_{k+a = m} E_\infty^{2j - \ell + k, a}.
%\label{hyper2inf}
%\fe
The Dolbeault cohomology groups ${\rm H}^a ( \cQ, \Lambda^j \cW (n) )$ can be computed using Borel-Weil-Bott theory.  The non-vanishing ones are
\ie
& {\rm H}^0({\cal Q}, \Lambda^0{\cal W}(n)) = [0,0,0,0,n],~~~~n\geq 0,
\\
& {\rm H}^{10}({\cal Q},\Lambda^0{\cal W}(n)) = [0,0,0,-8-n,0],~~~~ n\leq -8,
\\
& {\rm H}^0({\cal Q},\Lambda^1{\cal W}(n)) = [1,0,0,0,n],~~~~n\geq 0,
\\
& {\rm H}^{10}({\cal Q},\Lambda^1{\cal W}(n)) = [0,0,0,-9-n,1],~~~~ n\leq -9,
\\
& {\rm H}^0({\cal Q},\Lambda^2{\cal W}(n)) = [0,1,0,0,n],~~~~ n\geq 0,
\\
& {\rm H}^9({\cal Q},\Lambda^2{\cal W}(-8)) = [0,0,0,0,0],
\\
& {\rm H}^{10}({\cal Q},\Lambda^2{\cal W}(n)) = [0,0,1,-10-n,0],~~~~ n\leq -10,
\\
& {\rm H}^0({\cal Q},\Lambda^3{\cal W}(n)) = [0,0,1,0,n],~~~~ n\geq 0,
\\
& {\rm H}^1({\cal Q},\Lambda^3{\cal W}(-2)) = [0,0,0,0,0],
\\
& {\rm H}^{10}({\cal Q},\Lambda^3{\cal W}(n)) = [0,1,0,-10-n,0],~~~~ n\leq -10,
\\
& {\rm H}^0({\cal Q},\Lambda^4{\cal W}(n)) = [0,0,0,1,n+1],~~~~ n\geq -1,
\\
& {\rm H}^{10}({\cal Q},\Lambda^4{\cal W}(n)) = [1,0,0,-10-n,0],~~~~ n\leq -10,
\\
& {\rm H}^0({\cal Q},\Lambda^5{\cal W}(n)) = [0,0,0,0,n+2],~~~~ n\geq -2,
\\
& {\rm H}^{10}({\cal Q},\Lambda^5{\cal W}(n)) = [0,0,0,-10-n,0],~~~~ n\leq -10.
\label{bwblist2}
\fe
%, and the results are summarized in Appendix~\ref{sec:bwb}.  For 0D and 10D, we get the hypercohomology directly from (\ref{hyper2dol}).  In general dimensions, the hypercohomology is given by (\ref{hyper2inf}).  From results of Borel-Weil-Bott theory (\ref{bwblist2}),
We see that $E_1^{n + k, a} \neq 0$ only for $0 \leq k \leq 5, ~ a = 0, 10$, and for $(n, k, a) = (-2, 3, 1), (-8, 2, 9)$.  The following is a schematic diagram of the first page:
\begin{diagram}[2em]
E_1^{n+5, 0} & \quad & & \quad & \dotsb & \quad & & \quad & E_1^{n+5, 10} \\
\uTo^{d_1} & \luTo^{d_2}(2, 4) & & \quad & & \quad & & \quad & \uTo^{d_1} \\
E_1^{n+4, 0} & \quad & & \quad & \dotsb & \quad & & \quad & E_1^{n+4, 10} \\
\uTo^{d_1} & \quad & & \quad & & \quad & & \quad & \uTo_{d_1} \\
E_1^{n+3, 0} & \quad & E_1^{1, 1} ~ (n = -2) & \quad & \dotsb & \quad & & \quad & E_1^{n+3, 10} \\
\uTo^{d_1} & \quad & & \quad & & \quad & & \quad & \uTo^{d_1} \\
E_1^{n+2, 0} & \quad & & \quad & \dotsb & \quad & E_1^{-6, 9} ~ (n = -8) & \quad & E_1^{n+2, 10} \\
\uTo^{d_1} & \quad & & \quad & & \quad & & \luTo_{d_2}(2, 4) & \uTo^{d_1} \\
E_1^{n+1, 0} & \quad & & \quad & \dotsb & \quad & & \quad & E_1^{n+1, 10} \\
\uTo^{d_1} & \quad & & \quad & & \quad & & \quad & \uTo_{d_1} \\
E_1^{n, 0} & \quad & & \quad & \dotsb & \quad & & \quad & E_1^{n, 10}
\end{diagram}
The spectral sequence stabilizes on the third page for $n = -2, -8$, and on the second page otherwise.

For $n = -2, -8$, the $d_1$ map is trivial, so the second page is identical to the first page.
%Then we need to study the $d_2$ cohomology to go to the third page.
For $n = -2$, all chains on the second page are trivial except for
\ie
E_2^{1, 1} = {\rm Sym}^{j-3} (L^2)_d \to E_2^{3, 0} = {\rm Sym}^{j-5} (L^2)_d,
\fe
and for $n = -8$, all are trivial but for
\ie
E_2^{-8, 10} = {\rm Sym}^j (L^2)_d \to E_2^{-6, 9} = {\rm Sym}^{j-2} (L^2)_d.
\fe
{\it We will assume that the $d_2$ map is surjective in both cases.}  Then we are left with
\ie
& E_3^{1, 1} = ({\rm Sym}^{j-3} (L^2)_d)_{traceless} \quad j \geq 3, \quad (n, k, a) = (-2, 3, 1), \\
& E_3^{-8, 10} = ({\rm Sym}^j (L^2)_d)_{traceless} \quad j \geq 0, \quad (n, k, a) = (-8, 0, 10).
\label{3rdpage}
\fe
If this assumption fails, then there are additional classes in ${\bf H}^2$ and ${\bf H}^3$.

Next consider $n \neq -2, -8$.  The second page is given by the $Q$-cohomology of the following chain complexes
\ie
\label{0thCX}
0 & \to {\rm Sym}^j(L^2)_d\otimes[0,0,0,0,n]\to {\rm Sym}^{j-1}(L^2)_d\otimes [1,0,0,0,n] \\
& \to {\rm Sym}^{j-2}(L^2)_d\otimes [0,1,0,0,n] \to {\rm Sym}^{j-3}(L^2)_d\otimes [0,0,1,0,n] \\
& \to {\rm Sym}^{j-4}(L^2)_d\otimes [0,0,0,1,n+1]\to {\rm Sym}^{j-5}(L^2)_d\otimes[0,0,0,0,n+2]\to 0
\fe
and
\ie
\label{10thCX}
0 & \to {\rm Sym}^j(L^2)_d\otimes[0,0,0,-8-n,0]\to {\rm Sym}^{j-1}(L^2)_d\otimes [0,0,0,-9-n,1]
\\
&\to {\rm Sym}^{j-2}(L^2)_d\otimes [0,0,1,-10-n,0]\to {\rm Sym}^{j-3}(L^2)_d\otimes [0,1,0,-10-n,0]
\\
&\to {\rm Sym}^{j-4}(L^2)_d\otimes [1,0,0,-10-n,0]\to {\rm Sym}^{j-5}(L^2)_d\otimes[0,0,0,-10-n,0] \to 0.
\fe
%power of $Sym^* (L^2)_d$
Here a cochain vanishes if the number of copies of $(L^2)_d$ in the symmetric tensor product is negative or a Dynkin label is negative.  Even without knowing how $Q$ acts, the mere fact that $Q$ is $SO(10)$ equivariant can already lead us to conclude that certain representations must be in the $Q$-cohomology.
% We will only keep such ``conclusive'' classes (necessarily existent and whose location on the second page is known).  We illustrate our procedure by
Consider the following scenerios:
\begin{enumerate}
\item  If an irrep $r$ appears in the chain complex as $0 \to r \to 0$, then $r$ must be in the $Q$-cohomology.
\item  If $r$ appears as $0 \to r \to 3r \to r \to 0$, then we know that at least one copy of $r$ is in the $Q$-cohomology located at the middle.  {\it We will assume that this copy of $r$ is all there is in the $Q$-cohomology.}
\item  For $0 \to r \to r \to r \to 0$, we know that there must be one copy of $r$ in the $Q$-cohomology, but we do not know whether it is located on the left or on the right.  Further analysis is required.

%We do not keep this copy, and we know that we are definitely missing some classes in the hypercohomology. \\
\end{enumerate}

We restrict our attention to classes that preserve Lorentz or R-symmetry, and compute the $Q$-cohomology up to $j = 10$.  For the $a = 0$ chain complex (\ref{0thCX}), we consider $n$ ranging from $-2$ to $2j + 2$; for $a = 10$ (\ref{10thCX}), we consider $n$ from $-2j - 12$ to $-8$.  {\it We assume that no $Q$-cohomology appears outside our range of consideration. }

Scenerio~3 only appears in 2D, and only for representations that preserve Lorentz and break R-symmetry.  They are listed below, labelled by their $(j, n, a)$ values and representation of $SO(8)$.
\begin{itemize}
\item  {\bf $\bf (3, 2, 0)$ in $\bf [0, 1, 0, 0]$.}  The chain complex restricted to this representation is
\ie
r \to r \to r \to 0 \to 0 \to 0.
\fe
A $Q$-cohomology class at $k = 0$ will be in $\bf H^2$, while one at $k = 2$ will be in $\bf H^4$.  Since $n = 2j - \ell = 2 > -8$, there is an isomorphism ${\rm H}^* \cong {\bf H}^*$, and therefore we can determine which hypercohomology group contains $r$ by directly studying ${\rm H}^*$.  Consider ${\rm H}^2 ( {\rm Sym}^3 (YM_2) )_4 \cong {\rm H}^2 (N^3_c)_4$, where $N^j \equiv {\rm Sym}^j (YM_2)$.  Classes in ${\rm H}^2 (N^3_c)_4$ take the form $\lambda^2 D^3$, and there is only 1 copy of $[0, 1, 0, 0]$ in this tensor product, which is
\ie
(\lambda \Gamma^{01mnp} \lambda) D_p \circ D^2.
\fe
This expression is not $Q$-closed, so we conclude that $r$ is in ${\bf H}^4$ not ${\bf H}^2$.
\item  {\bf $\bf (4, 2, 0)$ in $\bf [0, 0, 1, 1]$.}  Similar to the previous case, classes in ${\rm H}^2 (N^4_c)_6$ takes the form $\lambda^2 D^4$, and there are 2 copies of $[0, 0, 1, 1]$ in this tensor product, which are
\ie
(\lambda \Gamma^{01mnp} \lambda) D^2 \circ D^2, \quad (\lambda \Gamma^{01q[mn} \lambda) D^{p]} \circ D_q \circ D^2.
\fe
No combination of the two is $Q$-closed, so we conclude that $r$ is in ${\bf H}^4$ not ${\bf H}^2$.
\item  {\bf $\bf (j \geq 5, 0, 0)$ in $\bf [j - 5, 0, 1, 1]$.}  They appear in the chain complex
\ie
0 \to 0 \to r \to 2r \to 3r \to r
\fe
Again there must be one copy of $r$ in either ${\bf H}^2$ or ${\bf H}^4$.  {\it We will assume that it is in ${\bf H}^4$. } \\
\end{itemize}

\noindent{\bf Validity of assumptions}

Because of the number of assumptions introduced above, the hypercohomology classes we find will naively be a subset of all the hypercohomology classes.  However, there are reasons to believe that such is not the case.
% we can be missing a lot of classes in the hypercohomology.
%In some sense, we are keeping the minimal set   The reasons to believe that such is not the case are as follows.
%First, Scenerio~3 never occurs in our computation, so there is no direct evidence showing that we are missing something.
First, in 0D there is an alternative way of computing the hypercohomology, which makes use of the quasi-isomorphism $\cW^* \otimes \cO(2) \hookrightarrow {\cal YM}$, and gives definite results.  The results there coincide with the results we obtain.  Second, consider (\ref{susyhyper})
\ie
{\rm H}_{1+2j} ({\bf susy}_d, \bC)_\ell \cong ({\rm coker}\,B_L/{\rm im}\,\iota_1) \bigoplus_{i+2j' = 1+2j}  {\bf H}^{2-i} (\cQ, {\rm Sym}^{j'} (\mathcal{YM}_d))_\ell.
\fe
In Section~\ref{Classification}, we will see that, our results for the right hand side already saturates the left hand side, so there is no room for missing classes in ${\bf H}^n$ for $n$ odd and $\leq 1$.  Our classification of exceptional D-terms is therefore rigorous. \\

\noindent{\bf Results}

We now present the results, organized by whether the hypercohomology classes preserve Lorentz or R-symmetry, or both.  The LiE program~\cite{lie} is used to facilitate this computation.  The classes in ${\bf H}^2 (\cQ, {\rm Sym}^j (\mathcal{YM}_d))_\ell$ are listed in Table~\ref{tab:hyper2}.  The classes in ${\bf H}^{1-2i} (\cQ, {\rm Sym}^j (\mathcal{YM}_d))_\ell$ for $i \geq 0, ~ \ell \geq 8$ in $d \geq 6$ are listed in Table~\ref{tab:hyperi}.  The rest are not needed for the purpose of classification.

\begin{table}[H]
\centering
\begin{tabular}{|c|c|c|c|}
\hline
$d$ & Lorentz $+$ R & Lorentz $-$ R & R $-$ Lorentz \\\hline\hline
all & $(3, -2, 3, 1)$ && $(2, 0, 2, 0)$ \quad 2-form \\
& $(0, -8, 0, 10)$ && singlet in $d = 2$ \\\hline
$d \leq 8$ && $(j \geq 4, -2, 3, 1) \quad ({\rm Sym}^{j-3} (L^2)_d)_{traceless}$ & \\
&& $(j \geq 1, -8, 0, 10) \quad ({\rm Sym}^j (L^2)_d)_{traceless}$ & \\\hline
10 &&& $(0, 2, 0, 0) \quad [00002]$ \\
&&& $(1, 1, 1, 0) \quad [10001]$ \\
&&& $(1, -9, 1, 10) \quad [00001]$ \\
&&& $(2, -10, 2, 10) \quad [00100]$ \\
&&& $(3, -11, 3, 10) \quad [01010]$ \\
&&& $(4, -12, 4, 10) \quad [10020]$ \\
&&& $(5, -13, 5, 10) \quad [00030]$ \\\hline
9 &&& $(0, 2, 0, 0) \quad [0002]$ \\
&&& $(1, 1, 1, 0) \quad [1001]$ \\
&&& $(2, -10, 2, 10) \quad [0100]$ \\
&&& $(3, -11, 3, 10) \quad [1001]$ \\
&&& $(4, -12, 4, 10) \quad [0002]$ \\\hline
8 &&& $(2, -10, 2, 10) \quad [1000]$ \\\hline
7 & $(2, -10, 2, 10)$ && \\\hline
5 & $(0, 2, 0, 0)$ && \\\hline
4 & $(1, 2, 0, 0)$ & $(0, 2, 0, 0) \quad [100]$ & \\\hline
3 && $(0, 2, 0, 0) \quad [010]$ & \\
&& $(1, 2, 0, 0) \quad [100]$ & \\\hline
2 & $(2, 0, 2, 0)$ & $(0, 2, 0, 0)$ [0011] & \\
&& $(1, 2, 0, 0) \quad [0100]$ & \\\hline
% && $(j \geq 5, 0, 2, 0) \quad [j - 5, 011]$ (?) & \\\hline
1 && $(0, 2, 0, 0) \quad [0002]$ & \\
&& $(1, 2, 0, 0) \quad [0010]$ & \\
&& $(2, 1, 1, 0) \quad [0001]$ & \\\hline
0 && $(0, 2, 0, 0) \quad [00002]$ & \\
&& $(1, 2, 0, 0) \quad [00011]$ & \\
&& $(2, 2, 0, 0) \quad [00020]$ & \\\hline
\end{tabular}
\caption{Classes in ${\bf H}^2 (\cQ, {\rm Sym}^j (\mathcal{YM}_d))_\ell$.  The numbers in parantheses are $( j, n = 2j - \ell, k, a )$, and the numbers in brackets are Dynkin labels of the corresponding $SO(10-d)$ or $SO(d)$ irreps.}
\label{tab:hyper2}
\end{table}

\begin{table}[H]
\centering
\begin{tabular}{|c|c|c|c|}
\hline
$d$ & Lorentz + R & R $-$ Lorentz \\\hline\hline
10 && $(0, -9-2i, 0, 10) \quad [000,1+2i,0]$ \\
&& $(1, -10-2i, 1, 10) \quad [000,1+2i,1]$ \\
&& $(2, -11-2i, 2, 10) \quad [001,1+2i,0]$ \\
&& $(3, -12-2i, 3, 10) \quad [010,2+2i,0]$ \\
&& $(4, -13-2i, 4, 10) \quad [100,3+2i,0]$ \\
&& $(5, -14-2i, 5, 10) \quad [000,4+2i,0]$ \\\hline
9 && $(0, -9-2i, 0, 10) \quad [000,1+2i]$ \\
&& $(1, -10-2i, 1, 10) \quad [001,2i]$ \\
&& $(2, -11-2i, 2, 10) \quad [010,1+2i]$ \\
&& $(3, -12-2i, 3, 10) \quad [100,2+2i]$ \\
&& $(4, -13-2i, 4, 10) \quad [000,3+2i]$ \\\hline
8 && $(1, -10-2i, 1, 10) \quad [01,i,i]$ \\
&& $(3, -12-2i, 3, 10) \quad [00,1+i,1+i]$ \\\hline
7 && $(1, -10-2i, 1, 10) \quad [1,i,0]$ \\\hline
6 & $(1, -10, 1, 10)$ & $(1, -10-2i, 1, 10) \quad [i \geq 1,0,0]$ \\\hline
%6 & \\\hline
\end{tabular}
\caption{Classes in ${\bf H}^{1-2i} (\cQ, {\rm Sym}^j (\mathcal{YM}_d))_\ell$ for $i \geq 0, ~ \ell \geq 8$ in $d \geq 6$.  The numbers in the parantheses are $( j, n = 2j - \ell, k, a )$, and the numbers in brackets are Dynkin labels of the corresponding $SO(10-d)$ or $SO(d)$ irreps.}
\label{tab:hyperi}
\end{table}

\section{More details on the classification of infinitesimal deformations}
\label{Classification}

Throughout this section we adopt the shorthand notation $N^j = {\rm Sym}^j (YM_d)$ and $\cN^j = {\rm Sym}^j (\mathcal{YM}_d)$.

\subsection{F-term deformations}

%From the long exact sequence (\ref{les2})
%\ie
%\cdots \to {\rm H}_1(L, N^j)_{\ell - 8} \stackrel{\delta_2}{\to} {\rm H}^2 (L, N^j)_\ell \to {\bf H}^2 (\cQ, \cN^j_\ell) \stackrel{\iota_2}{\to} {\rm H}_0 (L, N^j)_{\ell - 8} \to \cdots
%\label{hyperles}
%\fe
%we see that classes in ${\bf H}^2 ( \cQ, \cN^j_\ell)$ that are annihilated by $\iota_2$ come from classes in ${\rm H}^2 (L, N^j)_\ell$ that do not have a preimage in ${\rm H}_1 (L, N^j)_{\ell - 8}$ under $\delta_2$.  Since $\delta_2$ can be viewed as a full superspace integral, these are not of the form $A \tr G$ and hence are candidates for ``exceptional deformations''.

As explained in Section~\ref{sec:inf-def}, F-term deformations are identified with classes in the cokernel of $\delta$ in ${\bf H}^2 (\cQ, \cN^j)_\ell$ that are not annihilated by $i^*$.  Since classes in the cokernel of $\delta$ are in one-to-one correspondence with the classes in ${\bf H}^2 (\cQ, \cN^j)_\ell$ that are annihilated by $\iota$, to classify the F-term deformations, we examine each class in ${\bf H}^2 (\cQ, \cN^j)_\ell$ as listed in Table~\ref{tab:hyper2} (we identify them by their $(j, n, k, a)$ values), and determine whether it gives rise to an F-term deformation.  The classes that do are highlighted in boxes, for which we construct the corresponding representative in ${\rm H}^2(N_c)_\ell$.  %We will also express the representatives as a symmetric product of one $Q$-closed term with $Q$-exact terms to make the $Q$-closedness manifest.

We omit classes with $j = 0$ or $\ell$ odd ($n$ odd in Table~\ref{tab:hyper2}), because the former are annihilated by $i^*$ (see Appendix~\ref{Kernel of i^*}) and the latter do not respect the boson/fermion $\bZ_2$ grading.  We also note that a class in ${\bf H}^2 (\cQ, N)_\ell$ with $n = 2j - \ell > -8$ must have a preimage in ${\rm H}^2 (L, N^j)_\ell$ since ${\rm H}_0 (L, N)_{\ell - 8 < 2j} \cong 0$.

%must be annihlated by $\iota$

%must correspond to a class in ${\rm H}^2 (L, N^j)_\ell$.}  This is clear from from the exact sequence (\ref{hyperles}) and the fact that ${\rm H}_0 (L, N^j)_{\ell - 8 < 2j} \cong 0$.

%\item  {\it Deformations that preserve R-symmetry must either exist in all dimensions or only 0D.}  Dimensional reduction does not 

%\item  {\it An exceptional deformation in any dimension can always be lifted to an exceptional deformation in 10D (breaking Lorentz symmetry in general).}  This follows from the inclusion $TYM \subseteq YM_d$.

%UNLESS THE LIFT CAN BE REMOVED BY A FIELD REDEFINITION.

%We will try not to argue by Fact 2 but instead show directly why classes in ${\bf H}^2 (\cQ, \cN^j_\ell)$ that do not satisfy the criteria do not give rise to supersymmetric deformations.

\subsubsection{Lorentz and R-symmetry invariant deformations}

%Consider the first column in Table~\ref{tab:hyper2}.

\begin{itemize}

\item  \boxed{\text{\bf ${\bf (3, -2, 3, 1)}$ in all dimensions.}}  These classes correspond to
\ie
%\cO &=
\la (\lambda \Gamma^m \chi) \circ (\lambda \Gamma^n \chi) \circ F_{mn} \ra = \la Q D^m \circ Q D^n \circ F_{mn} \ra
\fe
in ${\rm H}^2 (N^3_c)_8$, giving rise to the $\delta \cL_{16}$ deformation in~\cite{Movshev:2009ba}, which is the supersymmetric completion of the $\tr F^4$ deformation.

\item  {\bf ${\bf (0, -8, 0, 10)}$ in all dimensions.}
%A corresponding class in ${\rm H}^2 (N_c)$ must have $j = 0$ and $\ell = 8$, which is impossible since $\lambda^\A \lambda^\B$ has $\ell = -2$.  We conclude that this class in ${\bf H}^2 (\cQ, \cN)$ must map must nontrivially into ${\rm H}_0 (L, N)$ and hence does not give rise to a deformation.
These classes sit in
\ie
\cdots \to {\rm H}^2 (N^0_c)_8 \cong 0 \to {\bf H}^2 (\cQ, \cN^0)_8 \stackrel{\iota_2}{\to} {\rm H}_0 (L, N^0)_0 \cong \bC \to \cdots
\fe
Here $N^0 = \bC$.  The only possible element in ${\rm H}^2 (N^0_c)_\ell$ is $\la \lambda^\A \lambda^\B \ra$ with $\ell = -2$, so ${\rm H}^2 (N^0_c)_8 \cong 0$.  Therefore these classes do not give rise to F-term deformations.

\item  {\bf ${\bf (2, -10, 2, 10)}$ in 7D.}  We will consider this class when we discuss the $(2, -10, 2, 10)$ classes in $d \geq 8$.  This class does not give rise to a deformation.

%\item  {\bf ${\bf (0, 2, 0, 0)}$ in 5D.}  This class corresponds to a constant $\cO = \lambda \Gamma^{12345} \lambda \in {\rm H}^2 (N^0_c)$.  It is annihilated by $i^*$ and does not give rise to a supersymmetric deformation.

\item  {\bf ${\bf (1, 2, 0, 0)}$ in 4D.}  This class corresponds to $\la \lambda \Gamma^{1234a} \lambda D_a \ra \in {\rm H}^2 (N^1_c)$.  It is annihilated by $i^*$.

\end{itemize}

\subsubsection{Lorentz invariant but R-symmetry breaking deformations}

%Consider the second column in Table~\ref{tab:hyper2}.

\begin{itemize}

\item  \boxed{\text{\bf ${\bf (j \geq 4, -2, 3, 1)}$ in $\bf d \leq 8$.}}
%These classes sit in
%\ie
%\cdots \to {\rm H}^2 (N^j_c)_{2j+2} \to {\bf H}^2 (\cQ, \cN^j_{2j+2}) \stackrel{\iota_2}{\to} {\rm H}_0 (L, N^j)_{2j-6} \cong 0 \to \cdots
%\fe
%Clearly, ${\rm H}_0 (L, N^j)_\ell = 0$ unless $\ell \geq 2j$.  There must be corresponding classes in ${\rm H}^2 (N^j_c)_{2j+2}$ in the correct $SO(10-d)_R$ representations.  Indeed we find
These classes correspond to the traceless part of
%the $(j-3)$-symmetric traceless representation of $SO(10-d)$
\ie
%\cO_{(a_1 \dotsb a_{j'})} &=
& \la (\lambda \Gamma^m \chi) \circ (\lambda \Gamma^n \chi) \circ (\chi \circ \Gamma_{mn(a_1} \chi) \circ D_{a_2} \circ \dotsb \circ D_{a_{j'})} \ra \\
&= \la Q D^m \circ Q D^n \circ (\chi \circ \Gamma_{mn(a_1} \chi) \circ D_{a_2} \circ \dotsb \circ D_{a_{j'})} \ra
\fe
in ${\rm H}^2 (N^j_c)_{2j+2}$, where $j' = j - 3$.
%\ie
%\boxed{
%\cO_{(a_1 \dotsb a_{j'})} = (\lambda \Gamma^m \chi) \circ (\lambda \Gamma^n \chi) \circ (\chi \circ \Gamma_{mn(a_1} \chi) \circ D_{a_2} \circ \dotsb \circ D_{a_{j'})}
%}
%\fe

%For $j = 4$, we can explicitly construct these classes:
%\ie
%%\boxed{
%\cO_a &= (\lambda \Gamma^m \chi) \circ (\lambda \Gamma^n \chi) \circ (\chi \circ \Gamma_{mna} \chi) = Q D^m \circ Q D^n \circ (\chi \circ \Gamma_{mna} \chi).
%%}
%\fe
%%, where $j' = j-3$ and traceless conditions are imposed.
%For $j >4$ in $d = 0$, we also have an explicit expression
%\ie
%%\boxed{
%\cO_{(a_1 \dotsb a_{j'})} &= (\lambda \Gamma^m \chi) \circ (\lambda \Gamma^n \chi) \circ (\chi \circ \Gamma_{mn(a_1} \chi) \circ D_{a_2} \circ \dotsb \circ D_{a_{j'})},
%%}
%\fe
%where $j' = j - 3$.

\item  {\bf ${\bf (j \geq 1, -8, 0, 10)}$ in $\bf d \leq 8$.}  These classes are in the $j$-symmetric traceless representations of $SO(10-d)$ and sit in 
\ie
\cdots \to {\rm H}^2 (N^j_c)_{2j+8} \to {\bf H}^2 (\cQ, \cN^j)_{2j+8} \stackrel{\iota_2}{\to} {\rm H}_0 (L, N^j)_{2j} \stackrel{\delta_3}{\to} {\rm H}^3 (L, N^j)_{2j+8} \to \cdots
\fe
%and can map to $D_{(a_1} \circ \dotsb \circ D_{a_j)} \in H_0 (L, N)_{2j}$
The R-symmetry breaking part of ${\rm H}_0 (L, N^j)_{2j}$ is generated by $\la D_{a_1} \circ \dotsb \circ D_{a_j} \ra$, which is in the $j$-symmetric tensor representation of $SO(10-d)$.  We claim that the traceless component of $\la D_{a_1} \circ \dotsb \circ D_{a_j} \ra$, which we denote by $\cO_{(a_1 \dotsb a_j)}$, is annihilated by $\delta_3$.  This would mean that the symmetric traceless classes we found in ${\bf H}^2$ map nontrivially under $\iota_2$, and have no preimage in ${\rm H}^2$.  Consider the diagram
\begin{diagram}
{\rm H}_{0}(L, N^j)_{2j} && \rTo^{\delta_3} && {\rm H}^{3}(L, N^j)_{2j+8}
\\
\uTo^{i_*} &&&& \dTo^{i^*} 
\\
{\rm H}_{0}(YM, N^j)_{2j} & \rTo^{A_0} & {\rm H}_{0}(YM, N^j)_{2j} & \rTo^{P}_\cong & {\rm H}^{3}(YM, N^j)_{2j+8}
\end{diagram}
Since $i_*$ here is surjective, we can pull $\cO_{(a_1 \dotsb a_j)}$ down to ${\rm H}_{0}(YM, N^j)_{2j}$.  Then under $A_0$, it maps to a sum of commutators of $U(YM_d)$ elements,\footnote{
This is equivalent to the statement that $\tr \cO_{(a_1 \dotsb a_j)}$, which is a BPS operator in MSYM, is annihilated by the successive action of 16 supercharges.
}
i.e., maps to a trivial representative in ${\rm H}_{0}(YM, N^j)_{2j}$.  By the commutative property of the diagram, $\cO_{(a_1 \dotsb a_j)}$ must be annihilated by $i^* \circ \delta_3$.  Another check for the claim is the following.  If $\cO_{(a_1 \dotsb a_j)}$ is indeed in the image of $\iota_2$, then following the lines of reasoning that led to (\ref{susyhyper}), we know that there must be a $j$-symmetric traceless representation inside ${\rm H}_{2j} ({\bf susy}_d, \bC)_{2j}$.  This is consistent with Table~\ref{tab:susyhom}.

%The $\delta_3$ map acts as follows
%{\bf justify?}
%\ie
%\cO \mapsto \left( D_\A \wedge D_\B \wedge D_\C \to T_{(\A\B\C)}^{\A_1 \dotsb \A_{11}} [ D_{\A_1}, \dotsb, [D_{\A_{11}}, \cO] \dotsb ] \right).
%\fe
%In MSYM theories, products of scalars in the symmetric traceless representation of R-symmetry are BPS operators, so the traceless part of $D_{(a_1} \circ \dotsb \circ D_{a_j)}$ is annihilated by $\delta_3$ and has a preimage in ${\bf H}^2 (\cQ, \cN^j)_{2j+8}$.

\item  {\bf ${\bf (1, 2, 0, 0)}$ in $\bf d \leq 3$.}  Similar to the $(1, 2, 0, 0)$ class in 4D, these classes are annihilated by $i^*$.

%\item  {\bf ${\bf (2, 1, 1, 0)}$ in 1D.}  This class corresponds to $\cO^\A = \lambda^\A (\lambda \Gamma^a \chi) \circ D_a \in {\rm H}^2 (N^1_c)$, which has odd grading and is unphysical.

\item  \boxed{\text{\bf ${\bf (2, 2, 0, 0)}$ in 0D.}}  This class corresponds to
\ie
%\boxed{
%\cO^{abcde} &= 
& \la (\lambda \Gamma^{abcde} \lambda) D^2 - 10 D^{[a} \circ (\lambda \Gamma^{bcde]f} \lambda) D_f \ra = - \la (\Gamma^{abcde})^{\A\B} Q D_\A \circ Q D_\B \ra
%}
\fe
in ${\rm H}^2 (N^2_c)_2$.

%It is in the self-dual 5-form representation, and has grading 2 (mass dimension 1) higher than the undeformed equations of motion.  From this we know that it has a string theory interpretation as D-instantons probing $AdS_5 \times S^5$.  The IKKT Lagrangian is deformed by
%\ie
%F_{abcde} \tr X^a [X^b, X^c] [X^d, X^e],
%\fe
%where $F_{abcde}$ is the self-dual 5-form flux.

\end{itemize}

\subsubsection{Lorentz-breaking but R-symmetry invariant deformations}
\label{sec:R-L}

%Consider the third column in Table~\ref{tab:hyper2}.

\begin{itemize}

\item  \boxed{\text{\bf $\bf (2, 0, 2, 0)$ in all dimensions.}} These correspond to
\ie
%\cO^{mn}
& \la (\lambda \Gamma^{mnpqr} \lambda) (\chi \circ \Gamma_{pqr} \chi) \ra
= \la \lambda \Gamma^m \chi \circ \lambda \Gamma^n \chi = Q D^{m} \circ Q D^{n} \ra
\fe
in ${\rm H}^2 (N^2_c)_4$, giving noncommutative Yang-Mills deformations.

%It is an antisymmetric rank-2 tensor, and the corresponding deformation parameter has mass dimension $-2$.  According to (4.27) of 9908142, the bosonic part of the corresponding Lagrangian deformation is
%\ie
%\delta \cL = \theta^{ab} ( F_{ac} F_{cd} F_{db} + A_a F_{cd} \partial_b F_{cd} ).
%\fe

\item  {\bf $\bf (2, -10, 2, 10)$ in $\bf d \geq 8$.}  Let us include the class in $d = 7$ in this discussion.  These classes sit in
\ie
\cdots \to {\rm H}^2 (L, N^2)_{14} \to {\bf H}^2 (\cQ, \cN^2)_{14} \stackrel{\iota_2}{\to} {\rm H}_0 (L, N^2)_6 \to \cdots
\fe
We now argue that these classes are not annilated by $\iota$.  In 10D this follows from Lemma~58 in~\cite{Movshev:2005ei}.  In $7 \leq d  \leq 9$, consider the commutative diagram
\begin{diagram}[2em]
{\bf H}^2 (\cQ, {\rm Sym}^2 ({\cal TYM}) )_{14} & \rTo^{\iota_2^{(10)}} & {\rm H}_0 (L, {\rm Sym}^2 TYM)_6 \\
\dTo^{dr} & & \dTo^{DR} \\
{\bf H}^2 (\cQ, {\rm Sym}^2 ({\cal YM}_d) )_{14}  & \rTo^{\iota_2^{(d)}} & {\rm H}_0 (L, {\rm Sym}^2 YM_d)_6
\end{diagram}
where $DR$ (dimensional reduction) is the induced map of the inclusion $TYM \subset YM_d$.  In 10D, ${\rm H}_0 (L, {\rm Sym}^2 TYM)_6$ is generated by $\la \chi \circ \Gamma^{mnp} \chi \ra$.  Under $DR$, one can show that its projection to the $(d-7)$-form representation of $SO(10-d)$ survives while the rest are annihilated.  By the commutative property of the diagram and the fact that there is only one $(d-7)$-form in ${\bf H}^2 (\cQ, ({\rm Sym}^2 {\cal YM}_d)_{14} )$, these classes in $7 \leq d \leq 9$ are also not annihlated by $\iota$.

%Let $c^{d}$ denote the class in ${\bf H}^2 (\cQ, ({\rm Sym}^2 {\cal YM}_d)_{14} )$

%If there were a deformation in $7 \leq d \leq 9$, then it could be lifted to 10D where the R-symmetry is trivial.  This proves that there is also no deformation in $7 \leq d \leq 9$.

%WE ALSO NEED TO ARGUE THAT THE LIFT OF AN EXCEPTIONAL DEFORMATION IS STILL EXCEPTIONAL.

\item  {\bf $\bf (4, -12, 4, 10)$ in $\bf d \geq 9$.}  These classes sit in
\ie
\cdots \to {\rm H}^2 (L, N^4)_{20} \to {\bf H}^2 (\cQ, \cN^4)_{20} \stackrel{\iota_2}{\to} {\rm H}_0 (L, N^4)_{12} \to \cdots
\fe
By the same argument as in the previous case, these do not give rise to F-term deformations.

%By Lemma~58 in 06paper, $\iota_2$ is an embedding in 10D, and therefore there is no corresponding class in ${\rm H}^2 (L, N^4)_{22}$.  If there were a deformation in 9D, then it can be lifted to 10D since the R-symmetry is trivial in both dimensions.  This proves that there is also no deformation in 9D.

\end{itemize}

\subsection{Exceptional D-term deformations}
\label{sec:ex-D}

%By the commutative diagram (\ref{diagram}), classes in ${\rm coker}\, i_*$ that are not annihilated by $\delta_2$ give rise to deformations that are not of the form $A \tr G$ for $G \in YM_d$.  Nonetheless, they are of the form $A \tr G$ for $G \in YM$, so we call them ``exceptional $A \tr G$ deformations''.  To classify them, we need a few ingredients:

As explained in Section, exceptional D-term deformations are identified with classes in the cokernel of $i_*$ that do not lie in the image of $\iota$.  According to the discussion in Appendix~\ref{susyhom},
%In Appendix~\ref{susyhom}, the study of a spectral sequence shows that
%\ie
%({\rm coker}\, i_* / {\rm im}\, \iota_1) \subset {\rm H}_{n, \ell} ({\bf susy}_d, \bC) \subset {{}}\bigoplus_{i+2j = n} {\rm H}_i (L, Sym^j YM_d)_\ell.
%\fe
%Another crucial observation is that $\iota_i$ is an isomorphism for $i \leq 0$ and $\iota_1$ is an injection, which follows from  the long exact sequence (\ref{les}) and our explicit knowledge of ${\rm H}^0 (L, {\rm Sym}^j YM_d)_\ell$ and ${\rm H}^1 (L, {\rm Sym}^j YM_d)_\ell$ (Table~\ref{HLd}).
%With this information, we see from the exact sequence (\ref{les}) that $\iota_i$ is an isomorphism for $i \leq 0$ and $\iota_1$ is an injection.
in order to classify the exceptional D-term deformations, we simply take all the classes in ${\rm H}_{n, \ell} ({\bf susy}_d, \bC)$ with odd $n$ and even $\ell$, and subtract by the classes inside $\bigoplus_{i+2j=n} {\bf H}^{2-i} (\cQ, {\rm Sym}^j \mathcal{YM}_d)_{\ell+8}$.  The SUSY cohomology groups are listed in Table~\ref{tab:susyhom}.  The homology groups can be obtained via the isomorphism
\ie
{\rm H}^{\ell, n} \equiv {\rm H}^n ({\bf susy}, \bC)_\ell \cong {\rm H}_{\ell-n} ({\bf susy}, \bC)_\ell \equiv {\rm H}_{\ell-n, \ell},
\fe
which exchanges the chiral and antichiral spinor representations.  Below we list the deformations found from the above procedure.

%The hypercohomology groups can be computed using the machinery outlined in Section~\ref{Hypercohomology}.

\subsubsection{Lorentz and R-symmetry invariant deformations}

%Consider the first column of Table~\ref{tab:susyhom}.  %After subtracting off the hypercohomology classes, we find the following deformations.

%The candidates are ${\rm H}_{3, 4}$ and ${\rm H}_{7, 12}$ in 10D, ${\rm H}_{5, 8}$ in 8D, and ${\rm H}_{3, 4}$ in 6D.  The class in 6D is the image of the $(1, -10, 1, 10)$ class in ${\bf H}^1 (\cQ, {\cal YM}_d)_{12}$ under $\iota_1$, and hence does not give rise to a deformation.

\begin{itemize}

\item  \boxed{\text{\bf $\bf H_{3, 4}$ and $\bf H_{7, 12}$ in 10D.}}  These are the $\delta\cL_{20}$ and $\delta\cL_{28}$ deformations in~\cite{Movshev:2009ba}.

\item  \boxed{\text{\bf $\bf H_{5, 8}$ in 8D.}}  This is the $SO(8) \times SO(2)$ invariant obtained from dimensional reducing the Lorentz-breaking $\bf H_{5, 8}$ class in 10D.

%\item  {\bf $\bf H_{3, 4}$ in 6D.}  This class is precisely the image of $(1, -10, 1, 10)$ in ${\bf H}^1 (\cQ, {\cal YM}_d)_{12}$ under $\iota_1$, and hence does not give rise to a deformation.

\end{itemize}

\subsubsection{Lorentz invariant but R-symmetry breaking deformations}

%Consider the second column of Table~\ref{tab:susyhom}.
All these classes have even $n$ and do not give rise to physical deformations.

\subsubsection{Lorentz-breaking but R-symmetry invariant deformations}

%Consider the third column of Table~\ref{tab:susyhom}.  %After subtracting off the hypercohomology classes, we find the following deformations.

%
\begin{itemize}

\item  \boxed{\text{\bf $\bf H_{5, 8} = [0, 1, 0, 0, 0]$ in 10D and $\bf H_{5, 8} = [1, 0, 0, 0]$ in 9D.}}  In 10D, this 2-form corresponds to
\ie
%\cO_{mn} =
\la 14D_\A \otimes \chi^\A \circ F_{mn} - D_\A \otimes (\Gamma_{mnpq}\chi)^\A \circ F_{pq} \ra
\fe
in ${\rm H}_1(L, {\rm Sym}^2 (YM_{10}))_8$.  The 9D class is just obtained from the 10D class by dimensional reduction.

\item  \boxed{\text{\bf $\bf H_{9, 14} = [0, 0, 0, 2, 0]$ in 10D.}}
The corresponding class in $H_1 (L, {\rm Sym}^4 (YM_{10}) )_{14}$ should be of the form
\ie
%\cO_{\A\B} = 
\la D_\C \otimes \chi^3 \circ F \ra
%\Gamma^{mnpqr}_{\A\B} D_s \otimes (\chi \circ \Gamma_{mnp} \chi) \circ (\chi \circ \Gamma_{qrs} \chi) + D_\C \otimes \chi^3 \circ F,
\fe
%where the second term makes $\cO_{\A\B}$ closed (there are four copies of $(00020)$ in $D_\C \otimes \chi^3 \circ F$).
There are four copies of $[00020]$ in $D_\C \otimes \chi^3 \circ F$.  A linear combination of them makes $\cO_{\A\B}$ nontrivial in the $Q$-cohomology.

%\item  {\bf ${\rm H}_{5, 8} = [1, 0, 0, 0]$ in 9D.}

\end{itemize}

\bibliography{onshellref} 
\bibliographystyle{JHEP}
 
\end{document}